\documentclass[pra,superscriptaddress,twocolumn]{revtex4}

\usepackage[SquareTraceBrackets]{quantum}
\usepackage{graphicx,bm,natbib,upgreek,amsbsy,amsmath,mathrsfs,accents}
\usepackage[dvipsnames]{color}
\definecolor{myblue}{named}{MidnightBlue}
\definecolor{mygreen}{RGB}{0,120,0}
\usepackage[colorlinks=true,citecolor=myblue,linkcolor=myblue,urlcolor=myblue]{hyperref}
\usepackage[T1]{fontenc}
\usepackage{newtxtext,newtxmath}
\usepackage[scaled]{helvet}
\usepackage{tikz}
\usetikzlibrary{arrows,decorations.pathmorphing,backgrounds,positioning,fit,petri}
\usepackage{gensymb}				
\usepackage{xr}
\usepackage[caption=false]{subfig}

\DeclareMathAlphabet{\mathscrbf}{OMS}{mdugm}{b}{n} 

\graphicspath{{./Graphics/}}
\DeclareGraphicsExtensions{.pdf, .jpg, .eps, .svg}

\newcommand{\cev}[1]{\reflectbox{\ensuremath{\vec{\reflectbox{\ensuremath{#1}}}}}} 

\begin{document}

\title{Quantum Fisher information for general spatial deformations of quantum emitters}
\author{Jasminder S. Sidhu}
\email{jsidhu1@sheffield.ac.uk}
\author{Pieter Kok}%
\email{p.kok@sheffield.ac.uk}
\affiliation{%
Department of Physics and Astronomy, The University of Sheffield, Sheffield, S3 7RH, UK}
\date{\today}

\begin{abstract}
We present a framework for the detection and estimation of deformations applied to a grid of sources. Our formalism uses the Hamiltonian formulation of the quantum Fisher information matrix (\textsc{qfim}) as the figure of merit to quantify the amount of information we have on the deformation matrix. Quantum metrology for grid deformations provides an ideal testbed to examine multi-parameter estimations for arbitrarily parameterised channel evolutions with generally non-commuting Hermitian generators. We generalise the local generator of translations for deformation parameters to multi-parameter estimations and use it to explore how well different deformations can be detected and corrected for. This approach holds for any deformation. We explore the application of our theory to the set of affine geometry maps. Both the configuration of the grid and the properties of the sources help to maximise the sensitivity of the \textsc{qfim} to changes in the deformation parameters. 
For the non-multiplicative Hamiltonian parameterisations resulting from grid rotations about any chosen axis, oscillatory dependence of the \textsc{qfi} surfaces for a specific interplay between mutual source separation distances and grid configurations.
\end{abstract}

\maketitle

\section{Introduction}

\noindent
For many quantities in physics, direct measurements are not possible, either in principle or due to experimental limitations. This is particularly true for quantum mechanical systems where variables such as entanglement, purity, and phase-time do not have associated quantum observables. In these situations, the values of a parameter are often inferred from a set of indirect measurements of a different observable, or set of observables. This procedure is addressed in quantum estimation theory~\cite{Holevo2011, Helstrom1976}; the formalism underlying the study of quantum sensing and metrology, where the objective is to find the fundamental precision bounds of parameter estimates and the optimal measurement strategies saturating them. Early work in this field predominantly focused on single parameter estimations of unitary, multiplicative parameters, such as phase and time~\cite{Toth2014_JPA, Leibfried2004_S, Giovannetti2011_NP}, whose quantum enhanced limit has been shown to always be attainable~\cite{Paris2009_IJQI}. However, applications of metrology to microscopy, optical, electromagnetic, and, gravitational field imaging often demand multiple parameter estimations. This has seen a surge of recent work focused on yielding quantum enhanced sensing from simultaneous estimation of multiple parameters~\cite{Fujiwara1994_METR, asymptotic_theory_book_2005, Matsumoto2002_JPA, Monras2011_PRA, Humphreys2013_PRL, Genoni2013_PRA, Crowley2014_PRA, Vidrighin2014_NC, Yao2014_PRA, Knott2016_PRA, Baumgratz2016_PRL}. Multi-parameter quantum enhanced sensing has provided a novel paradigm for investigating the information processing capabilities of multipartite or multimode quantum correlated states and measurements.

The quantum Cram{\'e}r-Rao bound (\textsc{qcrb}) provides a fundamental lower bound on the covariance matrix of parameter estimates. Although attainability of this bound is in general not guaranteed~\cite{Helstrom1976}, addressing its saturation for multi-parameter protocols have provided useful insights~\cite{Ragy2016_PRA}. The \textsc{qcrb} is proportional to the inverse of the quantum Fisher information (\textsc{qfi}); an important figure of merit in quantum parameter estimation theory~\cite{Holevo2011, Jarzyna2015_NJP}. This bound relates the information obtained about a parameter from measurements outcomes to the parameter estimate uncertainty. The calculation of the \textsc{qfi} for any physical system becomes one of the central tasks in quantum metrology, although generally, this is difficult. One approach to determining the \textsc{qfi} is to use the symmetric logarithmic derivative \textsc{(sld)} operator~\cite{Braunstein1994_PRL}. This approach is particularly suited to unitary quantum metrology, but less so for noisy processes, where the calculation involves complex optimisation procedures~\cite{Escher2011_N, Sarovar2006_JPA}. To address this, an extended Hilbert space approach may be taken where information about the parameter is obtained by observing both the system and its environment~\cite{Escher2012_PRL}. This method prescribes the \textsc{qfi} in terms of the state evolving Hamiltonian, and is well suited to many physical implementations of parameter estimations, including open quantum systems~\cite{Chin2012_PRL, Kolodynski2013_NJP, Alipour2014_PRL, Dobrzanski2014_PRL}. 

Previous work using the Hamiltonian parameter estimation approach has been dominated by phase-shift Hamiltonians; a special case where the parameter to estimate multiplies a parameter-independent Hermitian generator~\cite{Braunstein1996_AoP, Holevo1978_RMP}. This has proved particularly important in phase measurements, and, quantum enhanced Hamiltonian tomography, where unknown coefficients of the Hamiltonian are estimated after a suitable decomposition~\cite{Yurke1986_PRA, Sanders1995_PRL, Dorner2009_PRL, Skotiniotis2015_NJP}. For single parameter phase-shift Hamiltonians, the optimal probe state is an equal superposition of the eigenstates corresponding to the minimum and maximum eigenvalues of the Hamiltonian~\cite{Giovannetti2011_NP}. Recently, general Hamiltonians have been considered to estimate the profile of time-varying fields~\cite{Tsang2011_PRL, Magesan2013_PRA} and in gradient magnetometry~\cite{Urizar2013_PRA}. 
We note that in these applications either the Hamiltonian approach was not used to determine the \textsc{qfi}, or the problem constrained the generator to multiplicative parameterisations. Arbitrary Hamiltonian parameterisations would allow application of quantum metrology to a wider variety of quantum systems. General Hamiltonian parameterisations have been considered only recently by a few authors~\cite{Brody2013_E, Fraisse2017_PRA, Pang2014_PRA, Pang2016E_PRA, Seveso2017_arxiv}.

In this paper, we generalise the form of the generator derived in~\cite{Pang2014_PRA} for multi-parameter quantum metrology and obtain a general \textsc{qfim} for arbitrary unitary channel evolutions. We derive the form of the \textsc{qfim} for any grid deformation and explore the set of affine maps~\cite{Kadianakis2016_MMS}, including composite stretches, shears, and, rotations. Since the \textsc{qfim} depends only on the properties of the probe state and the configuration of the emitters, we explore how we can modify both to enhance our estimation sensitivity to determine the applied grid deformation.   

This paper is structured as follows: We introduce parameter estimation in quantum metrology in Sec.~\ref{sec:theory_qfi}. This will lead to the introduction of the quantum Fisher information (\textsc{qfi}). In Sec.~\ref{sec:general_formalism} we generalise the derivation of the Hamiltonian for arbitrary parameterisations, first presented by Pang and Brun~\cite{Pang2014_PRA}. Although the generator is Hermitian in this case, in general it is not. This is a well know problem in quantum theory and has wider consequences for deriving generalised uncertainty relations for a pair of operators in terms their variances~\cite{Luo2000_LMP, Robertson1929_PR}. The framework that statistical quantum estimation theory provides is best suited to deal with these problems~\cite{Holevo1978_RMP}. We derive the conditions to ascribe a self-adjoint operator to the physical quantity. In Sec~\ref{sec:grid_deformations}, we apply the formalism to the problem of grid deformations and understand how well we can detect arbitrary deformations by using the \textsc{qfi} as a metric. Physically motivated, we declare the best arrangement of sources that one should use to enhance sensitivity of detection for a set number of sources. Conclusions are presented in Sec.~\ref{sec:conclusions}.




\section{Multiparameter quantum Cram{\'e}r-Rao bound and attainability}
\label{sec:theory_qfi}
\begin{figure}[t!]
\begin{center}
\includegraphics[width =0.95\columnwidth]{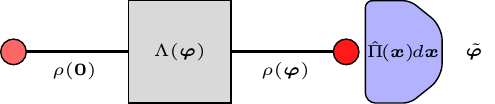}
\caption{(Colour online). General channel parameter estimation scheme. The quantum channel $\smash{\Lambda(\bm{\varphi})}$ parameterises an input probe state with the vector of parameters that we wish to estimate. The probe is measured by an operator of the form $\smash{\hat{\Pi}(\bm{x})d\bm{x}}$ and distributes estimates according to Born's rule. By data processing the measurement outcomes, we obtain our estimate $\smash{\tilde{\bm{\varphi}}}$. The goal of quantum metrology is in finding both the optimal probe state and observable that minimises the covariance matrix of unbiased estimates.}
\label{pic:estimation_scheme}
\end{center}
\end{figure}%

\noindent
We consider the estimation of parameters governed by the Hamiltonian $\smash{\hat{H}(\bm{\varphi})}$. The vector of parameters is $\smash{\bm{\varphi} = (\varphi_1, \ldots, \varphi_D)}$, with $\varphi_j \in B$, a Borel set and $j$ being the set of integers ranging between 1 and $D$ inclusive
, appear as arbitrary parameterisations of generally non-commuting Hamiltonians $\smash{\hat{H}_j}$, which describes the evolution of the $j\text{th}$ parameter. We note that the majority of earlier work have limited parameter estimates to coefficients in Hamiltonians proportional to $\smash{\hat{H}(\bm{\varphi}) = \sum_j\varphi_j\hat{H}_j}$. Although this has found significant uses in phase estimations and enhanced Hamiltonian tomography, we lift this limitation in what follows. 

The archetypal schema for Hamiltonian parameter estimations is illustrated in Fig.~\ref{pic:estimation_scheme}. We observe that the only access to the parameters $\bm{\varphi}$ is through probe measurements. The goal of optimal parameter estimation then requires a choice of an optimal input state and an optimal estimation scheme that maximises our information on $\bm{\varphi}$. Consider some fixed initial quantum probe state, $\smash{\rho(\bm{0})}$, is sent through a quantum channel, $\Lambda(\bm{\varphi})$, that describes the interrogated system and emerges parameterised by $\bm{\varphi}$. The emerging probe has the form $\smash{\rho(\bm{\varphi}) = \Lambda[\rho(\bm{0})](\bm{\varphi})}$ and is measured by some complete, self-adjoint observable $\smash{\hat{X} = \int d\bm{x} \, \bm{x}\hat{\Pi}(\bm{x})}$. The measurement outcomes $\bm{x}$ are distributed according to Born's rule $\smash{p(\bm{x}\vert\bm{\varphi}) = \tr{\rho(\bm{\varphi})\hat{\Pi}(\bm{x})}}$. Let $\bm{x}_1, \ldots, \bm{x}_\nu$ denote the measurement results of $\nu$ independent and identically distributed (\textsc{iid}) copies of the state $\smash{\rho(\bm{\varphi})}$, then an estimate $\tilde{\bm{\varphi}}$ of $\bm{\varphi}$ is achieved through an estimator that is a function of the measurement data. 
The performance of the estimator is quantified by the covariance matrix of the estimates $\bm{\Sigma}(\tilde{\bm{\varphi}})$. For some fixed positive operator valued measurement (\textsc{povm}), $\smash{\hat{\Pi}(\bm{x})d\bm{x}}$, the classical Cram{\'e}r-Rao bound (\textsc{ccrb}) states that the covariance matrix for any locally unbiased estimates of $\bm{\varphi}$ is lower bounded by the inverse of the classical Fisher information matrix (\textsc{cfim})~\cite{Kay1993}. That is,
\begin{align}
\nu\bm{\Sigma}(\tilde{\bm{\varphi}}) \geq \bm{\mathcal{F}}\left(\bm{\varphi}, \hat{\Pi}(\bm{x})\right)^{-1},
\label{eqn:mutivariate_ccrb}
\end{align} 
where $\nu$ is the number of repeated runs of the experiment and the \textsc{cfim} is defined by 
\begin{align}
\bm{\mathcal{F}}(\bm{\varphi}) = \int d\bm{x}\frac{\{\nabla_{\bm{\varphi}}p(\bm{x}\vert\bm{\varphi})\}\{\nabla_{\bm{\varphi}}p(\bm{x}\vert\bm{\varphi})\}^\top}{p(\bm{x}\vert\bm{\varphi})},
\label{eqn:cfim}
\end{align} 
with $\smash{\nabla_{\bm{\varphi}} = (\partial/\partial\varphi_1, \ldots, \partial/\partial\varphi_D)^\top}$. The \textsc{ccrb} is a matrix inequality which implies $\smash{\bm{\Sigma}(\tilde{\bm{\varphi}}) - \bm{\mathcal{I}}(\bm{\varphi})^{-1}}$ is positive semi-definite. If we introduce a positive definite, symmetric, and, real cost matrix, $\bm{R}$, a more general form of Eq~\eqref{eqn:mutivariate_ccrb} is
\begin{align}
\nu\Tr{\bm{R}\,\bm{\Sigma}(\tilde{\bm{\varphi}})} \geq \Tr{\bm{R}\,\mathcal{F}(\bm{\varphi})^{-1}}.
\label{eqn:weight_mutivariate_ccrb}
\end{align}
This inequality allows varying prioritisation for different parameters in $\bm{\varphi}$ by selecting a choice of $\smash{\hat{R}}$. This is a more meaningful inequality than Eq.~\eqref{eqn:mutivariate_ccrb} for multiparameter estimation protocols, where the objective can be one of minimising the weighted sum of variances of different parameters~\cite{Ragy2016_PRA} or alternatively, another weighting. Further, the cost matrix preserves dimensional consistency for the estimation of parameters with different units. The choice of $\smash{\bm{R}}$ results in a variety of precision bounds and estimators. It is chosen to minimise the Bayes risk, $\smash{\mathcal{R}}$ , defined as the expectation of the cost function~\cite{Kay1993}. The \textsc{ccrb} is a statement of the limitation of classical data processing. Our ability to resolve a set of parameters is constrained in their covariances by a quantity that depends only on the probability distribution from which we sample data. In the following discussion we explore how quantum estimation strategies can potentially provide greater advantages. Since the number of experimental repeats, $\nu$, provides a purely classical advantage, it will be suppressed hereafter.

Any change to the system parameters $\bm{\varphi}$, in the probe state, is captured by the multivariate probability distribution $\smash{p(\bm{x}\vert\bm{\varphi})}$. For some defined input probe state, and assuming that any self-adjoint observable $\smash{\hat{X}}$ can be measured, the goal of quantum estimation theory becomes one of finding an optimal measurement that minimises the estimate variances through changes to this distribution~\footnote{A second optimisation over input probe states is also required to optimally minimise the estimate variances. The pure states, the optimal probe is that which maximises the variance of the generator~\cite{Giovannetti2006_PRL, Giovannetti2011_NP}.}. Maximising the \textsc{cfi} over all measurement \textsc{povm}s yields the quantum Fisher information (\textsc{qfi}) which is a real, positive definite, and symmetric matrix. In the Braunstein and Caves formulation~\cite{Braunstein1994_PRL}, the \textsc{qfi} is given by  
\begin{align}
\bm{\mathcal{I}}(\bm{\varphi}) = \text{Re}\left(\Tr{\rho(\bm{\varphi})\bm{\mathcal{L}}\bm{\mathcal{L}}^\top}\right),
\label{eqn:qfim}
\end{align} 
where $\smash{\bm{\mathcal{L}} = (\mathcal{L}_1, \ldots, \mathcal{L}_D)^\top}$ is the Hermitian symmetric logarithmic derivative (\textsc{sld}) vector. The \textsc{sld} associated with the $j$th parameter is implicitly defined via
\begin{align}
2\partial_j\rho(\bm{\varphi}) = \left\{\rho(\bm{\varphi}), \mathcal{L}_j\right\}, 
\label{eqn:sld_implicit}
\end{align} 
where $\smash{\partial_j = \partial/\partial\varphi_j}$, and, $\{\hat{A},\hat{B}\} = \hat{A}\hat{B} + \hat{B}\hat{A}$ is the anticommutator. Given that the \textsc{qfi} is known, a generalisation of Eq.~\eqref{eqn:mutivariate_ccrb} leads to the multivariate quantum Cram{\'e}r-Rao bound (\textsc{qcrb}). Together, the multivariate bounds are summarised by
\begin{align}
\nu\Tr{\bm{R}\,\bm{\Sigma}(\tilde{\bm{\varphi}})} \geq \Tr{\bm{R}\,\mathcal{F}(\bm{\varphi})^{-1}} \geq \Tr{\bm{R}\,\mathcal{I}(\bm{\varphi})^{-1}}.
\label{eqn:mutivariate_bounds}
\end{align}
In addition to declaring fundamental bounds to estimate precisions allowed by quantum mechanics, quantum metrology addresses how these bounds can be approached. Saturating the \textsc{qcrb} requires equality in both the classical and quantum bounds in Eq.~\eqref{eqn:mutivariate_bounds}. Saturating the \textsc{ccrb} requires the use of an efficient, unbiased estimator that minimises the variance of parameter estimates~\cite{Braunstein1994_PRL}. For the single and multiple parameter cases, the \textsc{ccrb} is saturated in the asymptotic limit, $\nu \rightarrow \infty$, using the maximum likelihood estimator~\cite{Kay1993}. Developments pertaining the asymptotic saturability with finite $\nu$ have been investigated in the literature~\cite{Blandino2012_PRL, Braunstein1992_JPA}, although the performance of different probes changes the non-asymptotic regime~\cite{Rubio2017_JPC}. The performance of simultaneous metrology schemes are matched with separate schemes, and hence \emph{compatible}, if and only if there are no statistical correlations between estimators~\cite{Cox1987_POACI}. This ensures that off-diagonal elements of the covariance matrix of unbiased estimates reduces to zero, such that imperfect knowledge of one variable does not negatively impact the precision of estimating another~\cite{Ragy2016_PRA}.

Attainability of the \textsc{qcrb} is achieved for measurements that make the \textsc{cfi} equal to the \textsc{qfi}. Measurements that achieve this equality are referred to as \emph{optimal}. The single parameter \textsc{qcrb} is always saturable in the asymptotic limit, with optimal measurements given by the projection onto the eigenbasis of the \textsc{sld}; $\smash{\hat{X} = \bm{\mathcal{I}}^{-1}\bm{\mathcal{L}}}$~\cite{Braunstein1994_PRL, Paris2009_IJQI, Nielsen2000_JPA}. Implementation of this observable is generally difficult since it may depend on the exact values of the parameters $\bm{\varphi}$, which are assumed unknown. Adaptive measurement schemes that change the observable with increasing information on the parameters' true value have been suggested~\cite{Berry2000_PRL, Berry2002_PRA}. 

The multiparameter \textsc{qcrb} is generally not saturable. This is a consequence of the potential incompatibility of optimal measurements for each parameter. Asymptotic saturability of the bound is achieved if the \textsc{sld} associated with parameters $\varphi_j$ and $\varphi_k$ commute. This sufficiently proves the existence of an optimal measurement constructed from the common eigenbasis of the \textsc{sld}s. Even for non-commuting \textsc{sld}s, a weaker (sufficient and necessary) condition for asymptotically attainable \textsc{qcrb} is~\cite{Matsumoto2002_JPA, Ragy2016_PRA, Vaneph2013_QMQM} 
\begin{align} 
\Tr{\rho(\bm{\varphi})\left[\mathcal{L}_j, \mathcal{L}_k\right]}=0,
\label{eqn:qfim}
\end{align} 
where $[\hat{A},\hat{B}] = \hat{A}\hat{B} - \hat{B}\hat{A}$ is the commutator. Alternative figures of merit for precision estimation besides the \textsc{qcrb} have been used in multiparameter metrology~\cite{Ballester2004_PRA, Fujiwara1994_METR}. In this work, we use the \textsc{qcrb} metric in Eq.~\eqref{eqn:mutivariate_bounds}.

Different formulations of the \textsc{qfi} have been used in the literature. Braunstein and Caves introduced the superoperator \textsc{sld} to capture the dynamics of the state $\rho(\bm{\varphi})$. Although the \textsc{sld}s give the optimal measurements, their explicit expressions are generally difficult to obtain. This formalism is also not suited to performing the two-step optimisation of the \textsc{cfi} over optimal input probe states and optimal estimators for noisy processes~\cite{Escher2011_N}. Instead, the Kraus representation is more adept at describing general quantum channel operations~\cite{Kraus1983_Springer}. However, the Kraus decomposition of the quantum channel is non-unique. This causes the \textsc{qcrb} to be non-unique and proving the achievability of the bound is difficult~\cite{Sarovar2006_JPA}. In this work, we describe dynamics of the state through the system Hamiltonian. This formulation provides convenient attributes. First, the generator captures the dynamics of the parametrisation process of the state and is basis independent. This allows easy evaluation of the \textsc{qcrb} for different input probe states. Second, the \textsc{qfi} depends only on the generator and the initial states. Finally, entanglement between specific eigenstates of the generator can be used to construct an optimal state which maximises the \textsc{qfi}~\cite{Giovannetti2011_NP}. In the following section, we introduce this formalism and how to determine the generator of translations for the vector of parameters $\bm{\varphi}$.



\section{Generator formalism for quantum metrology}
\label{sec:general_formalism}

\noindent
In this section we review the generator formalism of the \textsc{qcrb}. We find that the \textsc{qfim} and hence the \textsc{qcrb} depends only the initial probe states and the generator of translations in the parameters $\bm{\varphi}$. A natural question that arises concerns the form of the generator. Hamiltonians which are parameterised by simple multiplicative factors have received much of the attention in this field~\cite{Holevo2011, Braunstein1994_PRL, Kolenderski2008_PRA, Vaneph2013_QMQM}. In this case, the generator of translations are parameter independent and are just the Hamiltonian. In this work, we consider more general forms of parameter-dependent generators due to its wider applicability in quantum parameter estimations. This arises when the parameters appear as different orders in the eigenvalues and/or eigenvectors decomposition of the Hamiltonian. This scenario is less developed, with a few authors considering general parameterisations of Hamiltonians. With this motivation, in subsection~\ref{subsec:multiparameter_general} we generalise the approach developed by Pang and Brun~\cite{Pang2014_PRA}, and Wilcox~\cite{Wilcox1967_JMP} to derive the form of the generator for arbitrary Hamiltonian parameterisations for multi-parameter estimations. Once the form of the generator is established, we are able to determine the multi-parameter \textsc{qfim}. How this is achieved is summarised in subsection~\ref{subsec:generator_qfi_matrix}.

\subsection{Multiparameter generators for arbitrary parameterisations}
\label{subsec:multiparameter_general}

\noindent
Consider evolving a quantum state $\rho(\bm{0})$ by some system Hamiltonian, $\smash{\hat{\mathcal{H}}(\bm{\varphi})}$. For the quantum channel illustrated in Fig.~\ref{pic:estimation_scheme}, the input probe state is evolved according to $\smash{\Lambda[\rho(\bm{0})] = \hat{U}(\bm{\varphi})\rho(\bm{0})\hat{U}^\dagger(\bm{\varphi})}$ with 
\begin{align}
\hat{U}(\bm{\varphi}) = \exp\left[-i\hat{\mathcal{H}}(\bm{\varphi})\right] \approx \exp\left[-i\hat{\bm{G}} \delta\bm{\varphi}^\top + \mathcal{O}(\bm{\varphi}^2)\right],
\label{eqn:unitary_form}
\end{align}
where we assume that we can linearise the unitary in $\smash{\bm{\varphi}}$. Our objective is to find the form of the generator $\smash{\hat{\bm{G}}}$ such that the unitary can be written approximately in the form shown in Eq.~\eqref{eqn:unitary_form}. The initial probe state is chosen so as to maximise its sensitivity to changes in $\bm{\varphi}$. Under infinitesimal changes to the parameterisation, this sensitivity may be characterised by the following Taylor expansion to second order of the evolving unitary 
\begin{align}
\hat{U}(\bm{\varphi} + \delta\bm{\varphi}) \approx \hat{U}(\bm{\varphi}) + \nabla_{\bm{\varphi}}\hat{U}(\bm{\varphi}) \delta\bm{\varphi}^\top + \frac{1}{2}\delta\bm{\varphi}\bm{H} \delta\bm{\varphi}^\top,
\label{eqn:unitary_multidim_Taylor}
\end{align}
where the parameter derivative vector $\bm{\nabla_{\bm{\varphi}}}$ and Hessian matrix $\bm{H}$ have the elements 
\begin{align}
\begin{split}
\left[\nabla_{\bm{\varphi}}\hat{U}(\bm{\varphi})\right]_j &= \partial_j\hat{U},\\
\left[\bm{H}(\bm{\varphi})\right]_{jk} &= \partial_j\left[\nabla \hat{U}(\bm{\varphi})\right]_k = \partial_{jk}^2\hat{U}(\bm{\varphi}),
\end{split}
\label{eqn:hessian_components}
\end{align}
with $\smash{\partial_j = \partial/\partial\varphi_j}$ and $\smash{\partial_{jk}^2 = \partial^2/\partial\varphi_j\partial\varphi_k}$. The parameter derivative vector differentiates the unitary with respect to each element of the parameter estimates $\bm{\varphi}$, which may generally correspond to different observables. This is a subtle, but important, distinction with the gradient vector $\bm{\nabla}$, which differentiates a function with different variables of the \emph{same} observable. To maintain full generality, we do not specify $\bm{\varphi}$ at this stage. Hence, we are unable to define the inner product of the operators in Eq.~\eqref{eqn:hessian_components} to determine their Hermitian conjugate. Hence we write
\begin{align}
\hat{U}(\bm{\varphi} + \delta\bm{\varphi})^\dagger \approx \hat{U}^\dagger(\bm{\varphi}) + \left(\nabla_{\bm{\varphi}}\hat{U}(\bm{\varphi})\right)^\dagger \delta\bm{\varphi}^\top + \frac{1}{2}\delta\bm{\varphi}\bm{H}^\dagger \delta\bm{\varphi}^\top,
\label{eqn:unitary_multidim_Taylor_hc}
\end{align}
For unbiased estimators, the data average in the asymptotic limit will give the true value $\smash{\bm{\varphi}}$. The general objective of finding a measurement and estimator with highest sensitivity to small variations in $\bm{\varphi}$ then justifies taking the approximation $\| \bm{R}\delta\bm{\varphi} \| \ll 1$. This implies that $\smash{\delta\bm{\varphi}\bm{H} \delta\bm{\varphi}^\top \approx 0}$, such that the unitary in Eq.~\eqref{eqn:unitary_multidim_Taylor} and Eq.~\eqref{eqn:unitary_multidim_Taylor_hc} may be truncated to first order in $\delta\bm{\varphi}$. By considering the sensitivities of the initial probe state to infinitesimal changes in $\bm{\varphi}$, we are able to find the generator of translations in $\bm{\varphi}$. Hence
\begin{align}
\begin{split}
\rho(\bm{\varphi} + \delta\bm{\varphi}) & = \left(\mathbbm{1} + \nabla\hat{U} \delta\bm{\varphi}^\top \hat{U}^\dagger\right) \rho(\bm{\varphi}) \left(\mathbbm{1} + \hat{U}\nabla\hat{U} \delta\bm{\varphi}^\top\right),\\
& \approx\exp\left[-i\hat{\bm{G}} \delta\bm{\varphi}^\top\right] \rho(\bm{\varphi})\exp\left[i\hat{\bm{G}}^\dagger \delta\bm{\varphi}^\top\right],
\end{split}
\label{eqn:evolved_state}
\end{align}
where elements of the local generator of parameter translations, $\smash{\hat{\bm{G}}}$, are given by
\begin{align}
\left[\hat{\bm{G}}(\bm{\varphi})\right]_j = \hat{G}_j = i[\partial_j\hat{U}(\bm{\varphi})]\hat{U}^\dagger(\bm{\varphi}).
\label{eqn:local_generator_general}
\end{align}
The derivative of the exponential operator, $\smash{\hat{U}}$, is given by Duhamel's formula
\begin{align}
\begin{split}
\partial_j\hat{U} &= -i\int_0^1 d\alpha \exp\left[-i(1-\alpha)\hat{\mathcal{H}}\right] \partial_j\hat{\mathcal{H}} \exp\left[-i\alpha\hat{\mathcal{H}}\right],\\
&= -i\hat{U}\hat{A}_j = -i\hat{A}_j\hat{U},
\label{eqn:operator_derivative_identity}
\end{split}
\end{align}
where $\smash{\hat{A}_j(\bm{\varphi})}$ is a Hermitian operator defined by the integral equation determined from Eq.~\eqref{eqn:operator_derivative_identity}. By combining Eq.~\eqref{eqn:operator_derivative_identity} with Eq.~\eqref{eqn:local_generator_integral_eq}, we find an integral equation representation for the generator of translations
\begin{align}
\hat{G}_j = \hat{A}_j = \int_0^1 d\alpha \exp\left[-i\alpha\hat{\mathcal{H}}\right] \partial_j\hat{\mathcal{H}} \exp\left[i\alpha\hat{\mathcal{H}}\right].
\label{eqn:local_generator_integral_eq}
\end{align}
A number of properties of the local generator immediately follow from this. First, it's Hermicity can be confirmed from Eq.~\eqref{eqn:local_generator_integral_eq} (Appendix~\ref{app:hermicity}) which is demanded for the state evolution expressed in Eq.~\eqref{eqn:evolved_state} to remain valid. Second, we notice that $\smash{\hat{\mathcal{H}}}$ is not necessarily equivalent to $\smash{\hat{\bm{G}}\bm{\varphi}^\top}$ in general as expected. An equality is only valid when the Hamiltonian has multiplicative dependence on the parameters $\bm{\varphi}$. Specifically, for Hamiltonian tomography $\smash{\hat{\mathcal{H}}(\bm{\varphi}) = \sum_j \varphi_j\hat{\mathcal{H}}_j}$, we recover $\hat{G}_j = \hat{\mathcal{H}}_j$. Third, despite our second observation, it is clear from Eq.~\eqref{eqn:operator_derivative_identity} that the generator of translations in $\varphi_j$ commutes with the unitary describing the channel evolution, $\smash{[\hat{G}_j, \hat{U}] = 0}$. This maintains conservation of the physical observable corresponding to the generator. Further, the generators of different parameters do not commute in general.

We turn our attention to the solution of the integral equation for $\smash{\hat{G}_j}$. We follow the method first proposed in~\cite{Wilcox1967_JMP}. This recasts the integrand $Y(\alpha)$ of Eq.~\eqref{eqn:local_generator_integral_eq} as a first order differential equation and introduces the super-operator $\smash{\hat{\mathscr{H}}Y(\alpha) = [\hat{\mathcal{H}}, Y(\alpha)]}$. Operationally, since this only depends on the operator $\smash{\hat{\mathcal{H}}}$, we are able to write the solution to Eq~\eqref{eqn:local_generator_integral_eq} in terms of its eigenvalues and eigenvectors. Hence, we assume that $\hat{\mathcal{H}}$ has $n_g$ unique eigenvalues, each with value $E_j$, $j\in[1, n_g]$ and degeneracy $d_j$ such that the corresponding eigenvectors are $\ket{\smash{E_j^{(k)}}}$, $k \in [1, d_j]$ satisfying $\smash{\braket{E_\alpha^{(\beta)}\vert E_\gamma^{(\delta)}} = \delta_{\alpha\gamma}\delta_{\beta\delta}}$. The Hamiltonian is then diagonal in this basis such that
\begin{align}
\hat{\mathcal{H}}(\bm{\varphi}) = \sum_{j=1}^{n_g} \sum_{k=1}^{d_j} E_j(\bm{\varphi})\ket{E_j^{(k)}(\bm{\varphi})}\bra{E_j^{(k)}(\bm{\varphi})},
\label{eqn:normal_hamiltonian}
\end{align}
where we allow both the eigenvalues and eigenvectors to depend on $\bm{\varphi}$. The resulting local generator of translations was solved in~\cite{Pang2014_PRA}, which we generalise for multiple parameters to give (Appendix~\ref{sec:generator_derivation})
\begin{align}
\begin{split}
\hat{G}_j(\bm{\varphi}) & = \sum_{k = 1}^{n_g} \partial_j E_k\hat{P}_k + 2\sum_{k\neq l}\sum_{m=1}^{d_k}\sum_{n=1}^{d_l} \exp\left[-i (E_k - E_l)/2\right] \\
& \times \sin\left[\frac{E_k - E_l}{2}\right] \Braket{E_l^{(n)}\middle\vert\partial_j E_k^{(m)}}\ket{E_k^{(m)}}\bra{E_l^{(n)}},
\end{split}
\label{eqn:final_generator_general_form}
\end{align}
where $\smash{\hat{P}_k = \sum_j\ket{\smash{E_k^{(j)}}}\bra{\smash{E_k^{(j)}}}}$ is the projector onto the $E_k$-eigenspace of $\hat{\mathcal{H}}(\bm{\varphi})$. The same result can be obtained by use of the Baker Campbell Hausdorff (\textsc{bch}) formulae on Eq.~\eqref{eqn:local_generator_integral_eq} (Appendix~\ref{sec:bch_approach_generator}). This approach represents the solution as an infinite series of nested commutators between $\hat{\mathcal{H}}$ and $\partial_j\hat{\mathcal{H}}$. From Eq.~\eqref{eqn:final_generator_general_form}, we observe that finding the generator is in general a difficult implicit problem since it requires the spectral decomposition of the Hamiltonian $\smash{\hat{\mathcal{H}}}$. Checking for consistency, we verify that for multiplicative factors we obtain the expected deduction $\smash{\hat{G}(\varphi) = \varphi\sum_{k}E_k\hat{P}_k = \varphi \hat{\mathcal{H}}}$, and, that the generator is Hermitian. This circumvents common methods, such as restricting the domain of the parameters $\bm{\varphi}$, or multiplying by the imaginary unit, to ensure Hermiticity or self-adjointness of operators. The form of the generator in Eq.~\eqref{eqn:final_generator_general_form} implies that the \textsc{qfi} can be separated into two parts. The first part is due to the dependence of the eigenvalues on $\varphi_j$, and the second due to the dependence of the eigenstates on $\varphi_j$. Increasing the channel \textsc{qfi} can be achieved by enhancing the sensitivity of the generator through additional terms in the Hamiltonian $\hat{\mathcal{H}}$~\cite{Fraisse2017_PRA}. Specifically, for time evolutions with parameter independent eigenvalues, Eq.~\eqref{eqn:final_generator_general_form} exhibits a periodic time dependence of the channel \textsc{qfi}~\cite{Pang2014_PRA}. Generally, this alone does not saturate the Heisenberg limit precision, but has been shown to with use of feedback controls~\cite{Yuan2015_PRL, Yuan2016_PRL}.


\subsection{Quantum Fisher information matrix}
\label{subsec:generator_qfi_matrix}

\noindent
In this subsection, we clarify the form of the \textsc{qfim} for multiparameters with generalised Hamiltonian evolutions. We start by considering the spectral decomposition of the evolved probe state
\begin{align}
\rho(\bm{\varphi}) = \sum_{j=1}^D \varrho_j(\bm{\varphi}) \ket{\varrho_j(\bm{\varphi})}\bra{\varrho_j(\bm{\varphi})},
\label{eqn:spectral_decomp}
\end{align}
where $D = \text{dim}[\text{supp}(\rho(\bm{\varphi}))]$ is the dimension of the support of $\rho(\bm{\varphi})$. We generalise the result for the \textsc{qfi} in~\cite{Liu2014_CTP} to multiple parameters
\begin{align}
\begin{split}
\left[\bm{\mathcal{I}}\right]_{mn} &= \hspace{-.2em}\sum_{j = 1}^D 4p_j \left[\text{Cov}\left(\cev{\partial}_m, \vec{\partial}_n\right)\right]_{j} \\
&- \hspace{-.2em}\sum_{j \neq k}^D \hspace{-.1em} \frac{8p_jp_k}{(p_j + p_k)}\Braket{\varrho_j\left\vert\cev{\partial}_m\right\vert\varrho_k} \Braket{\varrho_k \left\vert\vec{\partial}_n\right\vert\varrho_j}, 
\label{eqn:qfi_generator_form}
\end{split}
\end{align}
where $\{m,n \in \mathbbm{Z} \vert 1 \leq m, n \leq D\}$ define the elements of the \textsc{qfi} matrix elements, the arrows above the derivatives indicate the direction of operation, such that $\smash{(\vec{\partial}_j\ket{\varrho_k})^\dagger = \bra{\varrho_k}\cev{\partial}_j}$, and, the covariance matrix of the generators on the $j^\text{th}$-eigenstate of the initial state in Eq.~\eqref{eqn:qfi_generator_form} is defined as
\begin{align}
\begin{split}
\left[\text{Cov}\left(\cev{\partial}_m, \vec{\partial}_n\right)\right]_{j} & = \frac{1}{2} \Braket{\varrho_j\left\vert\left(\cev{\partial}_m \vec{\partial}_n + \cev{\partial}_n \vec{\partial}_m\right)\right\vert\varrho_j} \\
& - \Braket{\varrho_j\left\vert \cev{\partial}_m\right\vert\varrho_j}\Braket{\varrho_j\left\vert \vec{\partial}_n\right\vert\varrho_j}.
\label{eqn:covariance_matrix}
\end{split}
\end{align}
Despite unitary evolution of the probe, the \textsc{qfim} may depend on the parameters $\bm{\varphi}$, $\smash{\bm{\mathcal{I}} = \bm{\mathcal{I}}(\bm{\varphi})}$. We can easily re-write the \textsc{qfim} in terms of the generator by realising that $\smash{\ket{\smash{\partial_j\varrho(\bm{\varphi})}} \coloneqq \partial_j\hat{U}(\bm{\varphi})\ket{\varrho(\bm{0})} = -i\hat{G}_j\ket{\varrho(\bm{\varphi})}}$. Given the Hermicity of the generators, it suffices to replace both $\smash{\cev{\partial}_j}$ and $\smash{\vec{\partial}_j}$ with $\smash{\hat{G}_j}$ in Eq.~\eqref{eqn:qfi_generator_form} to obtain a general \textsc{qfim} for arbitrary unitary channel evolutions. This is the form of the \textsc{qfim} used in this work. We also observe that only for pure states, the \textsc{qfim} can be written as the covariance matrix of the generators
\begin{align}
\left[\bm{\mathcal{I}}\right]_{mn} = 4 \text{Re}\left[\Bra{\varrho}\hat{G}_m\hat{G}_n\Ket{\varrho} - \Bra{\varrho}\hat{G}_m\Ket{\varrho}\Bra{\varrho}\hat{G}_n\Ket{\varrho}\right].
\label{eqn:qfi_generator_pure_states}
\end{align}
Note that although the form of the \textsc{qfim} in Eq.~\eqref{eqn:qfi_generator_form} holds for probe states of arbitrary ranks, diagonalising a Hamiltonian of increasing rank is generally increasingly difficult.  

For completeness, we can intuitively relate the Hamiltonian formalism of the \textsc{qfim} written in Eq.~\eqref{eqn:qfi_generator_form} to the \textsc{sld} formalism. We note that from Eq.~\eqref{eqn:sld_implicit}, the \textsc{sld} describes the quantum dynamics of the system $\rho(\bm{\varphi})$. Unitary dynamics are generally given by the von-Neumann equation
\begin{align}
[\partial_j\rho(\bm{\varphi})]_{kl} = (\partial_j \varrho_k)\delta_{kl} + (\varrho_l - \varrho_k)\langle\varrho_k\vert\partial_j\varrho_l\rangle.
\label{eqn:von_neumann_eq}
\end{align}
where we defined $\smash{[\partial_j\rho(\bm{\varphi})]_{jk} \coloneqq \braket{\varrho_j \vert \partial_j\rho(\bm{\varphi})\vert\varrho_k}}$, used the fact that $\smash{\partial_j\braket{\varrho_k\vert\varrho_l}=0}$, and where we have dropped explicit dependence on $\bm{\varphi}$ on the \textsc{rhs}. Similarly, by decomposing the implicit definition of $\mathcal{L}$ in the eigenbasis of the state and combining with Eq.~\eqref{eqn:von_neumann_eq}, we obtain
\begin{align}
[\mathcal{L}_j]_{kl} = \frac{2(\partial_j \varrho_k)\delta_{kl}}{\varrho_k + \varrho_l} + \frac{2(\varrho_k - \varrho_l)\langle\partial_j\varrho_k\vert\varrho_l\rangle}{\varrho_k + \varrho_l},
\label{eqn:sld_final_form}
\end{align}
where $\ket{\smash{\partial_j\varrho(\bm{\varphi})}} \coloneqq \partial_j\hat{U}(\bm{\varphi})\ket{\varrho(\bm{0})}$. Note that for a pure state, this reduces to $\mathcal{L}_j=2\partial_j\rho(\bm{\varphi})$ as expected. Since $\smash{\hat{G}_j(\bm{\varphi}) \neq \hat{H}}$ for non-multiplicative parameterisations, we are prohibited us from writing the von-Neumann equation in Eq.~\eqref{eqn:von_neumann_eq} in a more recognisable manner. 

To conclude this section, we have provided a generalisation of the generator for multiparameter estimations. With this, the \textsc{qfim} can be used as a figure of merit from in Eq.~\eqref{eqn:qfi_generator_form} and the \textsc{sld} in Eq.~\eqref{eqn:sld_final_form} to question of saturability with optimal estimators. In the next subsection, we use the Hamiltonian formalism to determine how well we can detect deformations introduced to a grid of emitters. We also address how the grid configuration can be chosen to enhance the sensitivity of the \textsc{qfim}.


\section{Quantum metrology of grid deformations}
\label{sec:grid_deformations}

\noindent
In this section, we apply the Hamiltonian formalism introduced in the previous section to determine how well different spatial deformations of grids of photon emitters may be detected. One method of gauging how well deformations may be detected is to estimate the change in source positions by tracking changes to the emitted signature of the grid. We use the \textsc{qcrb} as the metric that describes the performance of our ability to estimate different deformations introduced to the grid and to define the ultimate theoretical precision bounds. This problem is isomorphic to source localisation which has been of considerable interest in the literature~\cite{Sidhu2017_PRA, Tsang2014_O, Fairhurst2011_CQG, Moerner2007_PNAS}. In the following, we work in the near-field regime, which defines the region of the electromagnetic field adjacent to the sources. This omits the need to propagate the emitted field to an alternative far-field region, allowing us to better concentrate on the source localisation problem.  
\begin{figure}[t!]
\centering
\includegraphics[width =\columnwidth]{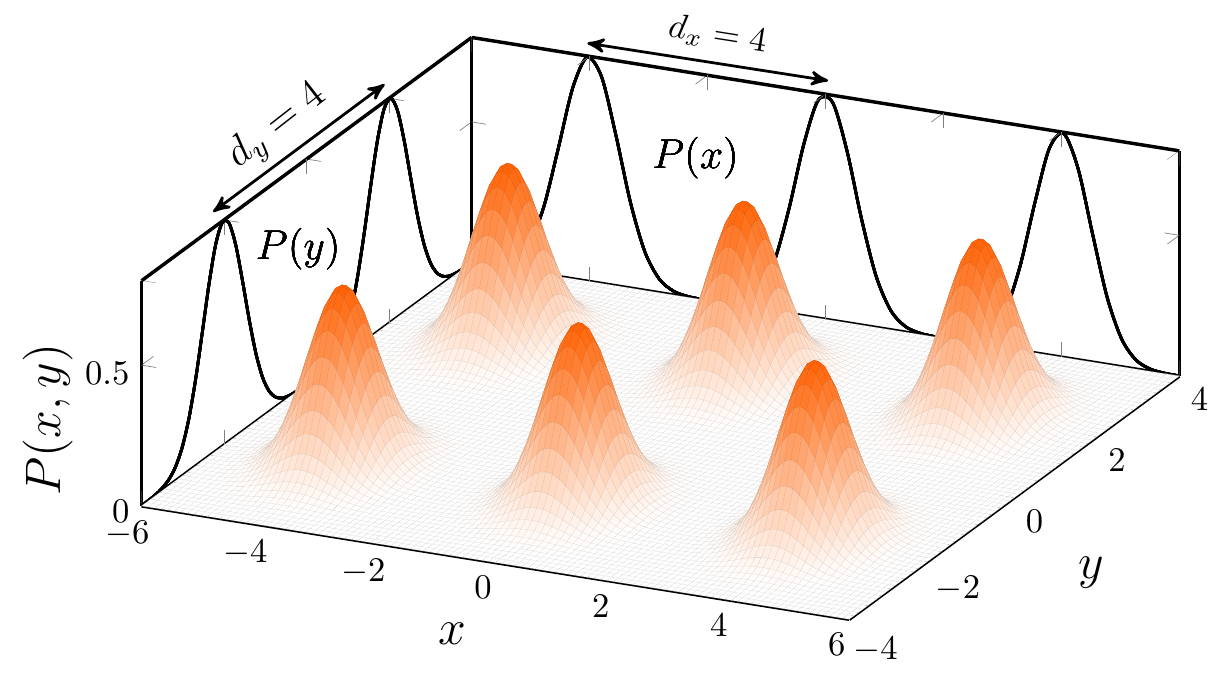}
\caption{A grid of $s = 6$ sources distributed in a $(3\times2)$ configuration, each with bi-variate Gaussian spatial profile, and source separation distances $d_x=4, d_y=4$ (arbitrary units). The projected probability distributions onto the $xz$ and $yz$ planes are illustrated by $P(x)$ and $P(y)$ respectively. We assume that the sources are independent and identically distributed (\textsc{iid}) such that covariance matrix of the emitted light in the near field is diagonal in the chosen basis. Any deformations introduced to this grid affects only the expected positions of the emitters whilst preserving the source covariances. This is demanded since any deformations should not change the emitters nature.}
\label{pic:stationary_spe_2d_grid}
\end{figure}%

Consider a two dimension grid of $s$ photon emitters arranged in a $(N\times M)$ configuration as illustrated in Fig~\ref{pic:stationary_spe_2d_grid}. Homogenous deformations introduced to the grid of sources will affect the expected mean positions of each source, not their covariances. By tracking these changes we hope to make an estimate of the type of deformation the grid has been subjected to.  The expected position of the $j$th source, $\smash{\bm{\mu}_j} = (\mu_{j_x}, \mu_{j_y})^\top$, $\smash{j = \{1, 2, \ldots, s\}}$, is chosen to be symmetric about the grid centre $O$ such that
\begin{align}
\begin{split}
\mu_{j_x} &= \left[\text{Mod}\left[j - 1,N\right] - \left(\frac{N-1}{2}\right)\right] d_x, \\
\mu_{j_y} &= \left[\left\lceil{\frac{j}{N}}\right\rceil - \left(\frac{M+1}{2}\right)\right] d_y,
\end{split}
\label{eq:grid_x_positions}
\end{align}
where $\text{Mod}[a,b]$, $\{a,b\} \in \mathbbm{R}$, defines the modulo operation that returns remainder of the division $a/b$, $\smash{\lceil{a}\rceil = \text{Ceiling}[a]}$ returns the smallest integer that is greater than or equal to $a$, $d_x$ and $d_y$ define the source separation distance in the $x$ and $y$ direction respectively. The convention chosen for labelling emitters $j$ is from bottom left to top right, with increasing $j$ running along the rows. We now assume that the $j$th-source emits $n_j$ photons, each with a bi-normal spatial distribution with mean position $\bm{\mu}_j$ and covariance matrix $\smash{\bm{\Sigma}_j}$. In what follows, we reserve bold typesetting for tuples. Then, the distribution of each emitted photon may be described by the first and second moments according to 
\begin{align}
f(\bm{r}_j, \bm{\mu}_j) = \frac{1}{2\pi\abs{\bm{\Sigma}_j}^{1/2}}\exp\left[-\frac{1}{2}(\bm{r}_j - \bm{\mu}_j)^\top\bm{\Sigma}_j(\bm{r}_j - \bm{\mu}_j)\right],
\label{eqn:bivariate_gaussian}
\end{align}
where 
\begin{align}
\bm{\Sigma}_j = \begin{pmatrix}
\sigma_{j_x}^2 & \rho\sigma_{j_x}\sigma_{j_y} \\
\rho\sigma_{j_x}\sigma_{j_y} & \sigma_{j_y}^2
\end{pmatrix}, \quad \bm{\mu}_j = \left(\mu_{j_x}, \mu_{j_y}\right)^\top,
\label{eqn:source_covariance_matrix}
\end{align}
and $\rho$ is the correlation coefficient between $x$ and $y$. The pure state describing the emitted light from the initial undeformed grid of $s$ \textsc{iid} sources in the near field may then be written as
\begin{align}
\ket{\Psi(\bm{0})} &= \int \, d\bm{R}^{\bm{n}} \; f(\bm{R}, \bm{M})^{\bm{n}/2} \hat{a}^\dagger(\bm{R})^{\bm{n}} \ket{\bm{0}},
\label{eqn:nphoton_state_untransformed}
\end{align}
where $\smash{d\bm{R} = \prod_{j=1}^s d\bm{r}_j^{n_j}}$, $\smash{f(\bm{R}, \bm{M})^{\bm{n}} = \prod_{j=1}^s f(\bm{r}_j, \bm{\mu}_j)^{n_j}}$ is the mode function with which the state is normalised, $\ket{\bm{0}} = \ket{0}^{\otimes s}$ is the global vacuum state (i.e., no excitations in any mode), and, $\smash{\hat{a}^{\dagger \bm{n}}(\bm{R}) = \prod_{j=1}^s \hat{a}_j^{\dagger n_j}(\bm{r}_j)}$ is the multimode creation operator composed of tensor products of creation operator on the Fock space of the $j$th source, $\smash{\hat{a}_j^\dagger(\bm{r}_j)}$. These creation operators do not have explicit time dependence. From its Heisenberg equation of motion, we may write the position dependent creation operator as 
\begin{align}
\hat{b}^\dagger_{j}(\bm{r}_j) = \int d\bm{k}_j \; \hat{a}_j^\dagger (\bm{k}_j) \exp\left[i\bm{r}_j\bm{k}_j\right],
\label{eqn:creation_operator_fourier}
\end{align}
Substituting Eq.~\eqref{eqn:creation_operator_fourier} into Eq.~\eqref{eqn:nphoton_state_untransformed} and using the definition of the mode function $f$, we write the state in the Fourier domain as  
\begin{align}
\ket{\Psi(\bm{0})} &= \int \, d\bm{K}^{\bm{n}} \; \varphi(\bm{K}, \bm{M})^{\bm{n}/2} \hat{a}^\dagger(\bm{K})^{\bm{n}} \ket{\bm{0}},
\label{eqn:nphoton_state_untransformed_fourier}
\end{align}
where the mode profile function $\varphi$ is
\begin{align}
\varphi(\bm{k}_j) = \sqrt{\frac{2\sigma_{j_x}^2\sigma_{j_y}^2}{\pi}}(1 - \rho)^{1/4}\exp\left[i\bm{k}_j^\top\bm{\mu}_j - \bm{k}_j\bm{\Sigma}_j\bm{k}_j\right],
\label{eqn:fourier_mode_function}
\end{align}
which differs from the characteristic function of $f$ owing to the square root of the binormal Gaussian distribution in the state definition in Eq.~\eqref{eqn:nphoton_state_untransformed}. Grid deformations will have the effect of parameterising the initial probe state $\rho(\bm{0})$ by changing the expected source positions $\smash{\bm{\mu}_j}$ to $\smash{\tilde{\bm{\mu}}_j}$. 

\begin{figure}
\subfloat[Composite stretching.]{\includegraphics[width=0.47\columnwidth]{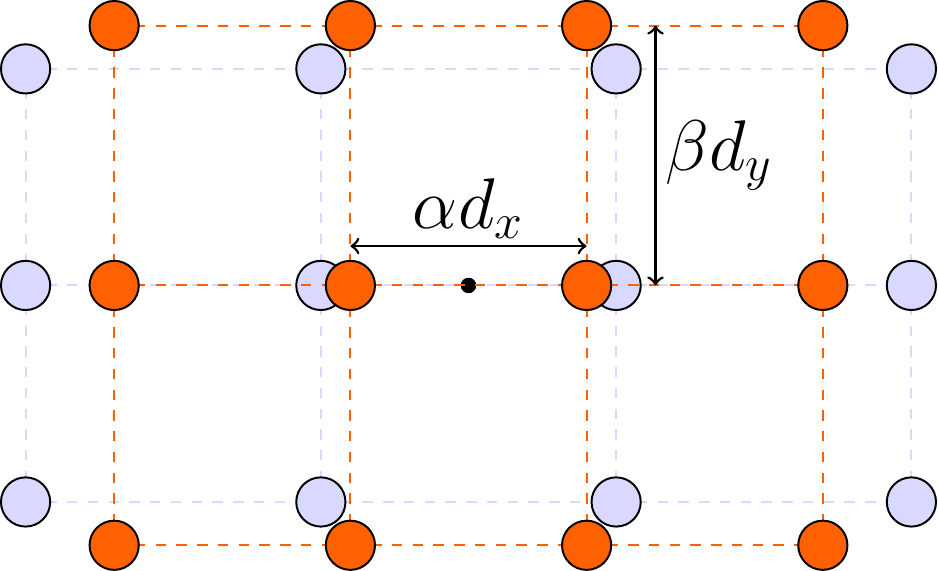}\label{fig:comp_stretch}} \hspace{8pt}
\subfloat[Rotation.]{\includegraphics[width=0.47\columnwidth]{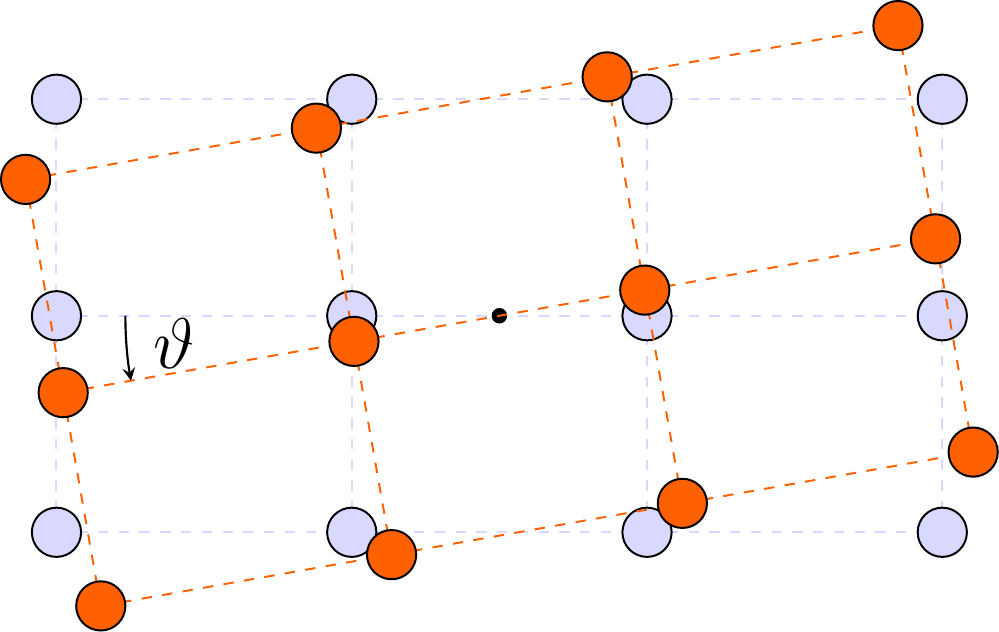} \label{fig:rotation}} \\
\subfloat[Composite shear.]{\includegraphics[width=0.47\columnwidth]{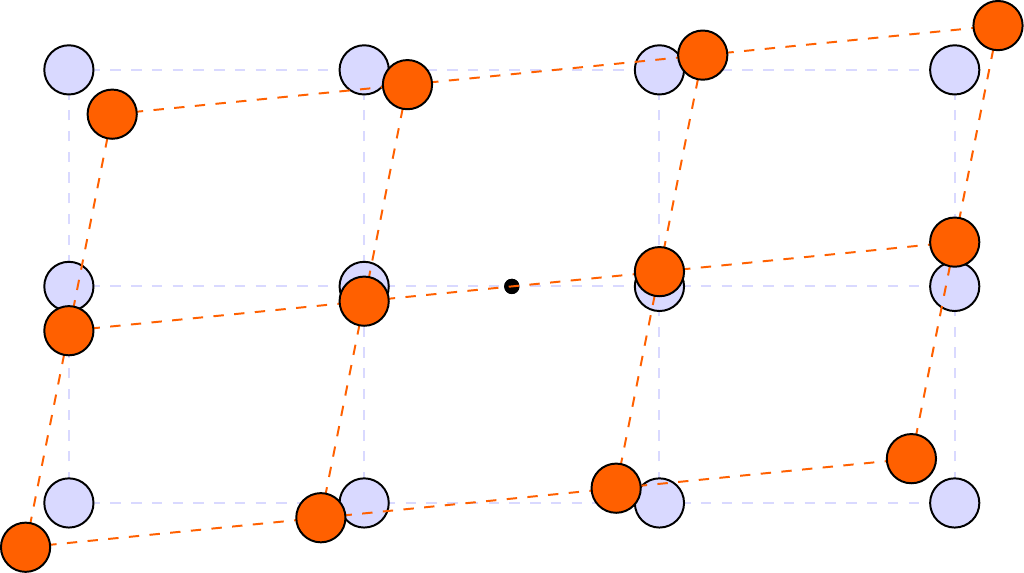} \label{fig:comp_shear}} \hspace{8pt}
\subfloat[Inhomogenous.]{\includegraphics[width=0.47\columnwidth]{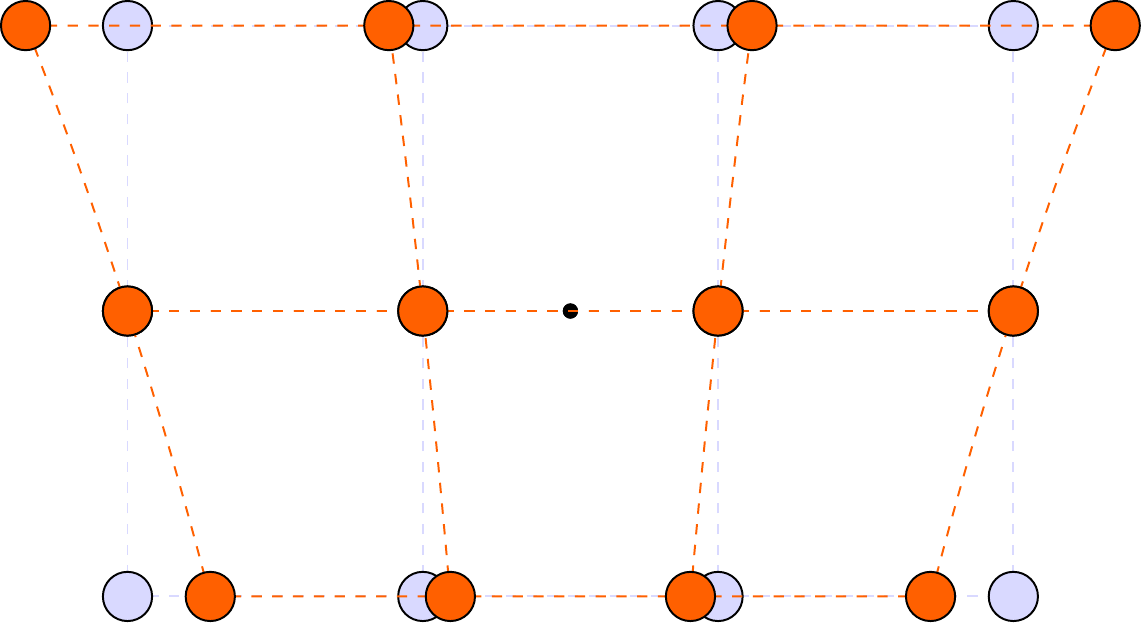} \label{fig:inhomogenous}}
\caption{(Colour online). This figure illustrates different deformations of the grid of Gaussian distributed sources shown in Fig~\ref{pic:stationary_spe_2d_grid}. The deformed grids are shown with orange sources, and compared with the original-undeformed grid shown with blue sources. The central black dot defines the grid centre $O$. In Fig.~\ref{fig:comp_stretch} we consider a composite stretching which provides a platform for multiparameter estimation of the unit-less multiplicative factors $\alpha$ and $\beta$. In Fig.~\ref{fig:rotation} we consider rotations about the grid centre $O$. Whilst this is a single parameter estimation, it is not a simple multiplicative factor. Fig.~\ref{fig:comp_shear} represents a grid shear, and in Fig.~\ref{fig:inhomogenous}, we consider a position dependent inhomogeneous (non-linear) deformation, where the x-coordinates are transformed to $\smash{x' = x\exp(y/\gamma)}$ for $\gamma\in\mathbbm{R}$.}
\label{pic:grid_deformation_types}
\end{figure}

The complexity of the calculation depends on how the grid deformation, $\bm{F}$, parameterises the generator. The case for multiplicative factors such as grid stretching, was considered in~\cite{Sidhu2017_PRA}. In that work, the \textsc{qfi} was used to estimate the source separation distance $\bm{d} = (d_x, d_y)$. The generator for $\bm{d}$ used there was
\begin{align}
\hat{\bm{G}} = -\sum_{j} \int d\bm{k}_j \bm{\Lambda}_j\bm{V}_j^\top \hat{n}(\bm{k}_j),
\label{eqn:2d_generator}
\end{align}
with $\smash{\bm{\Lambda}_j = \text{diag}(\mu_{j_x}, \mu_{j_y})}$, $\smash{\bm{V}_j = (\bm{F}^\top - 1)\bm{k}_j^\top}$, $\smash{\hat{n}}$ is the number operator, and, $\hat{\bm{G}} = (\hat{G}_{d_x}, \hat{G}_{d_y})^\top$ generates dynamics in the source separation distances. We note that this is a self-adjoint operator. For an array of equidistant sources, the generator becomes $\hat{G} = \xi \sum_{j=1}^N \mu_j'\int dk_j \, k_j \, \hat{n}_j(k_j)$ and this system has been explored metrologically in~\cite{Sidhu2017_PRA}. In this work, we consider applying a wider range of homogeneous and inhomogeneous deformations to a two dimensional grid of sources and evaluating their impact on the \textsc{qfi}. 
By determining how the \textsc{qfi} changes with different deformations, we can track changes to the source coordinates in the grid and apply corrective measures to negate the effects of the deformation. This would also enable evaluation of stresses and strains in materials when deformed. Fig~\ref{pic:grid_deformation_types} illustrates the types of grid deformations that we would like to consider in this section.

Recall that the form of the unitary that parameterises the position of a grid of $s$ sources in terms of some parameterisation is 
\begin{align}
\hat{U}(\bm{\varphi}) = \exp\left[- i \hat{\mathcal{H}}(\bm{\varphi})\right],
\label{eqn:general_parameter_unitary}
\end{align}
where, 
\begin{align}
\hat{\mathcal{H}}(\bm{\varphi}) = -\sum_{j=1}^s\int d\bm{k}_j \left(\bm{k}_j\bm{u}_j^\top(\bm{\varphi})\right) \hat{n}_j(\bm{k}_j).
\label{eqn:transformation_generator}
\end{align}
By inspection of this generator, the Fock-basis is suitable to span this operator. Defining the eigenvectors as 
\begin{align}
\ket{\bm{n}} \equiv \ket{n_1(\bm{k}_1), \ldots, n_s(\bm{k}_s)},
\label{eqn:sudogenerator_eigenvector}
\end{align}
then from the eigenvalue problem $\smash{\hat{\mathcal{H}}(\bm{\varphi})\ket{\bm{n}} = E(\bm{\varphi})\ket{\bm{n}}}$, we obtain
\begin{align}
E(\bm{\varphi}) = -\sum_{j=1}^s n_j \int d\bm{k}_j \left(\bm{k}_j\bm{u}_j^\top(\bm{\varphi})\right)
\label{eqn:sudogenerator_eigenvalue}
\end{align}
for the corresponding eigenvalues. Since the eigenvectors have no dependence on the parameter to be estimated, we have that the generator of translations in deformation $[\bm{\varphi}]_j = \varphi_j$ is 
\begin{align}
\hat{G}_j = \sum_{k = 1}^{n_g}\partial_j E_k\ket{\bm{n}}\bra{\bm{n}}.
\label{eqn:actual_generator}
\end{align}
This generator is Hermitian and has units of momentum. This is expected since the grid sources undergoes spatial translations according to some deformation $\bm{F}$. For any homogenous deformation, there is no net translation of the sources about the grid centre, $O$ (see Fig.~\ref{pic:grid_deformation_types}). This suggests that $\smash{\braket{\hat{G}_j} = 0}$, which can be confirmed through directly calculation. Hence, the \textsc{qfim} for a pure state becomes
\begin{align}
\left[\bm{\mathcal{I}}^Q\right]_{mn} = 2 \Braket{\hat{G}_m\hat{G}_n + \hat{G}_n\hat{G}_m} = 4 \Braket{\hat{G}_m\hat{G}_n}.
\label{eqn:qfi_generator_purestate}
\end{align}
The second equality on Eq.~\eqref{eqn:qfi_generator_purestate} used a further property of the generators: $\smash{[\hat{G}_m, \hat{G}_n]} = 0$, which is trivial to see from Eq.~\eqref{eqn:actual_generator}. Using the definition of the state $\smash{\ket{\Psi}}$, we can calculate the \textsc{qfi}, which yields
\begin{align}
\begin{split}
\left[\bm{\mathcal{I}}^Q\right]_{mn} &= \sum_{j=1}^s \frac{n_j^2}{(1 - \rho^2)} \left\{\frac{(\partial_m u_{j_x}) (\partial_n u_{j_x})}{\sigma_{j_x}^2} + \frac{(\partial_m u_{j_y})(\partial_n u_{j_y})}{\sigma_{j_y}^2} \right.\\
&- \left.\frac{\rho\left[(\partial_m u_{j_x})(\partial_n u_{j_y}) + (\partial_m u_{j_y})(\partial_n u_{j_x})\right]}{\sigma_{j_x}\sigma_{j_y}}\right\}.
\label{eqn:qfi_general_deformation}
\end{split}
\end{align}
We note that the \textsc{qfi} depends only on the properties of the probe state and the grid configuration. This suggests it may be possible to modify both properties to maximise the sensitivity of the \textsc{qfim} to changes in $\smash{\bm{\varphi}}$. We explore this possibility in the following material. The result detailed in Eq.~\eqref{eqn:qfi_general_deformation} is central to the paper since it completes the metrology approach for grid deformations. It is valid for all possible deformations. For non-homogenous deformations, the deforming matrix will also depend on the source index number $j$,
\begin{align}
\bm{u}_j^\top = \bm{F}_j\bm{\mu}_j^\top.
\label{eqn:general_deformation_shift}
\end{align}
This significantly increases the computation time so in the remainder of this paper, we limit our treatment to homogenous deformations. This is the set of affine deformations~\cite{Kadianakis2016_MMS, DeformationTheoryBook2010_Springer}, composed entirely of linear transformations such as rotations, shear, and, composite stretches. In this instance, the deformation matrix is the same for all emitters in the grid. 
To proceed further, we must consider specific parameterisations. In the following sub-sections, we consider different grid deformations and use the Hamiltonian approach to calculate the precision of detecting changes introduced to the grid. We first check that the formalism works for deformations with multiplicative parameterisations of the Hamiltonian, such as grid stretching and shearing, before considering the simplest non-multiplicative parameterisation resulting from grid rotations about any chosen axis.

\subsection{Grid stretching and shearing deformations}
\label{subsec:grid_stretches}
\begin{figure}
\subfloat[$\bm{\mathcal{I}}^Q_{\alpha\alpha}$]{\includegraphics[width=0.485\columnwidth]{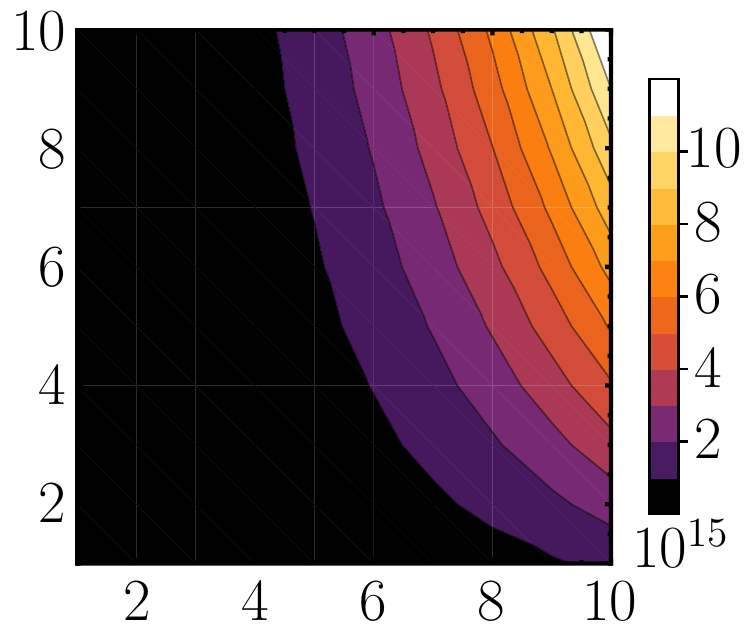}\label{fig:qfi_alpha_contour}} \hspace{3pt}
\subfloat[$\bm{\mathcal{I}}^Q_{\beta\beta}$.]{\includegraphics[width=0.485\columnwidth]{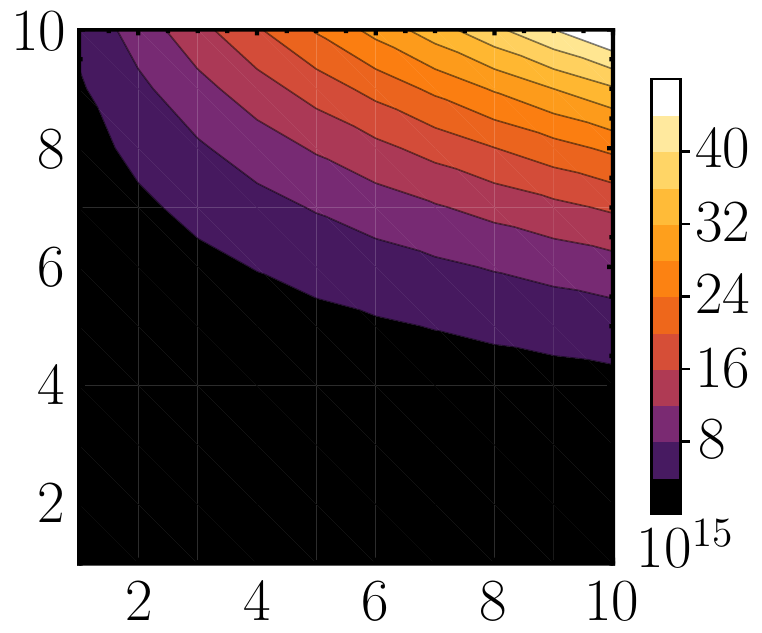} \label{fig:qfi_beta_contour}}
\caption{(Colour online). Diagonal elements of the \textsc{qfim} for a composite stretching deformation of the grid (illustrated in Fig~\ref{fig:comp_stretch}) as a function of the grid size $N$ ($x$-axis) and $M$ ($y$-axis), with $d_x=1$m, $d_y=2$m, $\rho = 0.5$, and $\sigma_{j_x} = \sigma_{j_y} = \sigma = 300$nm---typical of photons from quantum dots. The zero off-diagonal elements is a consequence of assuming \textsc{iid} sources along the grid. From Fig.~\ref{fig:qfi_alpha_contour}, we verify that the \textsc{qfi} for the stretching factor $\alpha$ along the x-axis increase as the grid size increases. Increasing the number of emitters along the stretching direction is favoured for this. This recovers the same finding in~\cite{Sidhu2017_PRA}. Fig.~\ref{fig:qfi_beta_contour} demonstrates similar conclusions for the \textsc{qfi} for $\beta$.}
\label{pic:stretch_QFI_contourplots}
\end{figure}%
%
%
%
%
%
%

\noindent
We start by first considering the simplest grid deformation. The stretching shown in Fig~\ref{fig:comp_stretch} stretches the grid by a factor $\alpha$ in the x-direction and factor $\beta$ in the y-direction. Each source in the grid is deformed according to the same deformation matrix described by 
\begin{align}
\bm{F}_{\text{stretch}} = \begin{pmatrix}
\alpha & 0 \\
0 & \beta
\end{pmatrix},
\label{eq:stretch_deformation_matric}
\end{align}
This is a multiplicative, bi-variate estimation scheme. We would like to determine how the \textsc{qfi} behaves with changing $\smash{\bm{\varphi} = (\alpha, \beta)}$. Using Eq.~\eqref{eqn:qfi_general_deformation} and Eq.~\eqref{eqn:general_deformation_shift}, the $(2\times2)$ \textsc{qfim} can be written as
\begin{align}
\left[\mathcal{I}^Q\right]_{kl} = \sum_{j=1}^s\frac{q_j^2 \mathcal{A}_{kl} \mu_{j_k}\mu_{j_l}}{\sigma_{j_k}\sigma_{j_l}},
\label{eq:qfi_grid_streching}
\end{align}
where $\smash{q_j}$ is the number of photons emitted by each source, $\smash{\mathcal{A}_{kl} = \delta_{kl}/(1-\rho^2) - \rho(1 - \delta_{kl})}$ with $k,l = \{\alpha, \beta\}$, $s$ is the total number of sources, and where we assumed each source has the same standard deviation in the ordered basis for the grid, $\smash{\sigma_{j_x} = \sigma_{j_y} = \sigma}$. We note that the only source dependence of the \textsc{qfim} arises from the expected mean positions of the emitters. All other terms detail the properties of continuous variable Gaussian states for each source, and factorise out of the summation since we assume each source is \textsc{iid}, and are subject to the same deformation $\smash{\bm{F}}$. This greatly simplifies the calculation of the \textsc{qfim}, whose diagonal elements are illustrated in Fig~\ref{pic:stretch_QFI_contourplots}. We observe that the \textsc{qfim} is independent of $\smash{\bm{\varphi}}$, which is entirely a consequence of the multiplicative parameterisation of the Hamiltonian $\smash{\hat{\mathcal{H}}}$. However, we are still able to maximise the sensitivity of the \textsc{qfi} by adjusting the grid configuration. To understand how, we provide an analytic expression for the diagonal elements
\begin{figure}[t!]
\begin{center}
\includegraphics[width =\columnwidth, height=0.6\columnwidth]{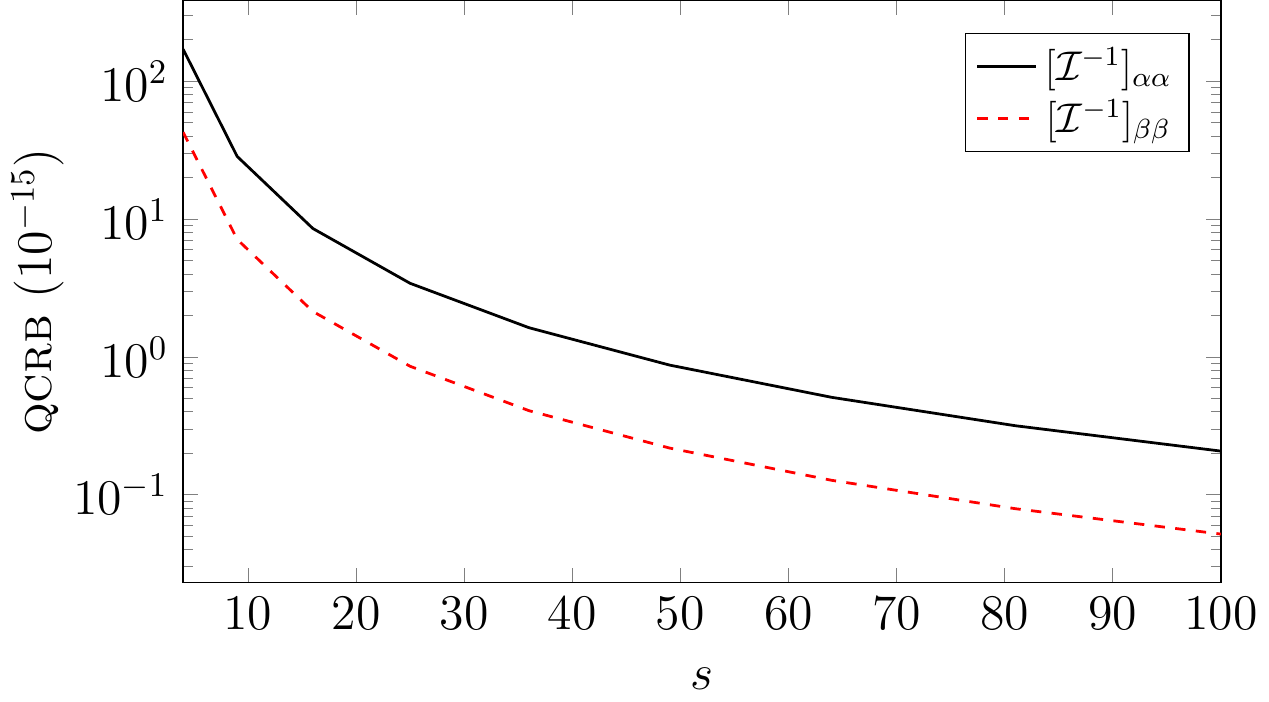}
\caption{(Colour online). \textsc{qcrb} along the $N=M$ slice of the contour plots in Fig.~\ref{pic:stretch_QFI_contourplots}. The black solid line illustrates the \textsc{qcrb} for $\alpha$ and the red dashed line for $\beta$. For a square grid, $N=M$, the relative difference between the estimate variances is accounted for by the fraction of source separations distances in the $x$ and $y$ directions, according to Eq.~\eqref{eqn:qfi_stretch_ratio}. Along this plane, we choose $d_x=1$, $d_y=2$ (and observe a constant factor 4 difference between the two lines).}
\label{pic:qfi_slice}
\end{center}
\end{figure}%

\begin{align}
\left[\bm{\mathcal{I}}\right]_{\alpha\alpha} = \frac{d_x^2s(N^2 -1)}{12\sigma^2(1-\rho^2)}, \quad \left[\bm{\mathcal{I}}\right]_{\beta\beta} = \frac{d_y^2s(M^2 -1)}{12\sigma^2(1-\rho^2)}.
\label{eqn:qfi_stretch_diagonal_analytic}
\end{align}
It is clear from this that maximising the sensitivity of the \textsc{qfi} is achieved by increasing the number of emitters along the same direction as the grid stretching is performed, for identical number of total sources, $s$. This is in agreement with results obtained in~\cite{Sidhu2017_PRA}. Further we note that the \textsc{qfi} for $\alpha$ is effectively just the mirror image of the \textsc{qfi} for $\beta$, with the mapping $N\leftrightarrow M$, which can also be observed from the contour plots in Fig.~\ref{pic:stretch_QFI_contourplots}. The ratio, $\mathscr{R}$, of the diagonal elements
\begin{equation}
\mathscr{R} = \left(\frac{d_x}{d_y}\right)^2 \frac{N^2 - 1}{M^2 - 1},
\label{eqn:qfi_stretch_ratio}
\end{equation}
provides a clear instructive guide for maximising either the \textsc{qfi} for $\alpha$ or for $\beta$, by controlling the grid configuration. To illustrate this, we plot the \textsc{qcrb} along the $N=M$ plane of the contour plots for the stretching deformation in Fig.~\ref{pic:qfi_slice}.

The results discussed for composite grid stretching can immediately be applied to grid shearing, since they both parameterise the Hamiltonian with similar multiplicative factors. The composite grid shear illustrated in Fig.~\ref{fig:comp_shear}, can be described by a shear in the horizontal direction with factor $\iota$ and factor $\kappa$ in the vertical direction. The deformation matrix for each source along the grid is then described by
\begin{align}
\bm{F}_{\text{shear}} = \begin{pmatrix}
1 & \iota \\
\kappa & 1
\end{pmatrix}.
\label{eq:shear_deformation_matric}
\end{align}
The generator of translations in $\smash{\bm{\varphi} = (\iota, \kappa)}$ has the same form as that for the composite stretching, except for an interchange of basis. The consequence of this is that the \textsc{qfi} for $\iota$ is exactly that shown in Fig.~\ref{fig:qfi_beta_contour}. Similarly, the \textsc{qfi} for $\kappa$ is that shown in Fig.~\ref{fig:qfi_alpha_contour}.



\subsection{Grid rotations}
\label{subsec:grid_stretches}

\begin{figure}[t!]
\begin{center}
\includegraphics[width =\columnwidth, height=0.6\columnwidth]{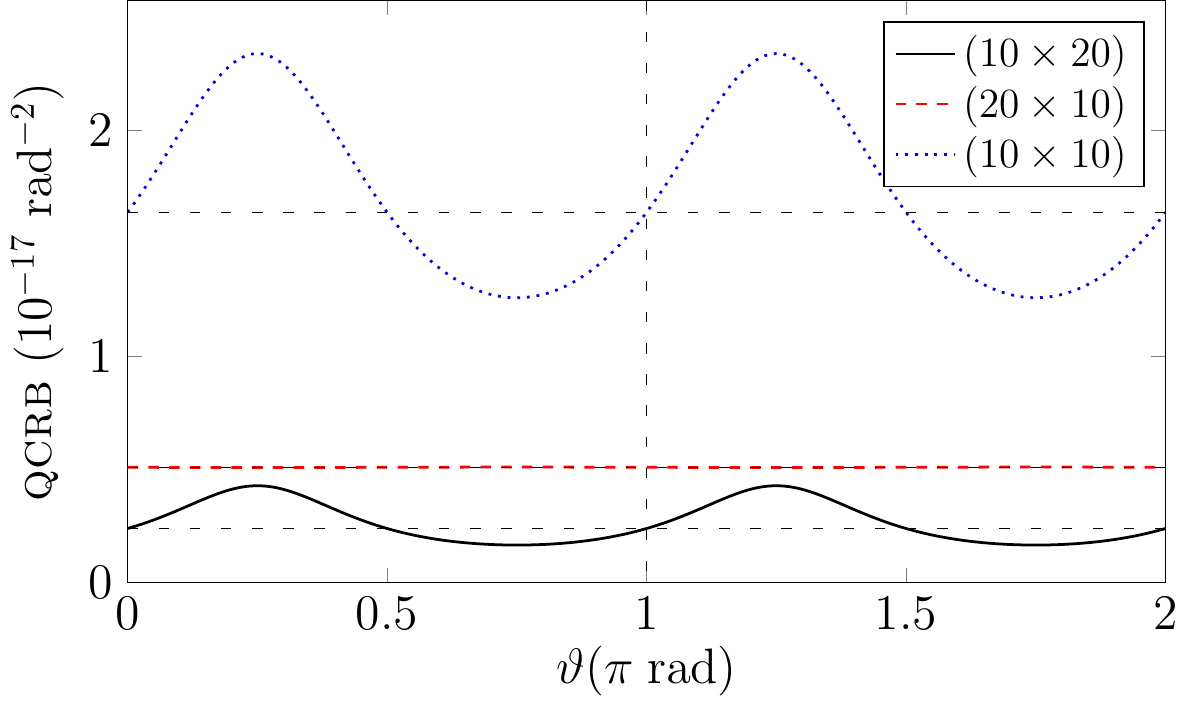}
\caption{(Colour online). The \textsc{qcrb} for a grid composed of $s=(n \times m)$ source configuration undergoing a rotation about the centre $O$, with $d_x=1$m, $d_y=2$m, correlation parameter $\rho = 0.5$, and, $\sigma_{j_x} = \sigma_{j_y} = \sigma = 300$nm. We observe a greater precision of detecting the rotation as the number of sources in the grid increases. The amplitude of the oscillatory behaviour can be made more pronounced by re-arranging sources such that $N>M$ for positive $\rho$, and $N<M$ for negative $\rho$. Further, we note that since rotations form a homogeneous deformation about the grid centre, a period of $\pi$ is observed in the variation of the \textsc{qcrb}.}
\label{pic:qcrb_grid_rotations}
\end{center}
\end{figure}

\noindent
In this subsection, we consider the rotation map illustrated in Fig.~\ref{fig:rotation}. For rotations in the counterclockwise direction, the expected position of each emitter transforms according to the deformation matrix
\begin{align}
\bm{F}_{\text{rot}} = \begin{pmatrix}
\cos(\vartheta) & -\sin(\vartheta) \\
\sin(\vartheta) & \cos(\vartheta)
\end{pmatrix}.
\label{eq:stretch_deformation_matrix}
\end{align}
The transformed expected emitter positions then satisfy the following properties: $\smash{\partial_\vartheta\tilde{u}_{j_x} = -u_{j_y}}$ and $\smash{\partial_\vartheta\tilde{u}_{j_y} = u_{j_x}}$. Estimating the rotation angle $\vartheta$ is a single parameter estimation protocol, only with a non-multiplicative factor of the Hamiltonian. Our expression for the \textsc{qfim} written in Eq.~\eqref{eqn:qfi_general_deformation} accounts for this. From the properties of the transformed source positions, we calculate the \textsc{qfi} to be 
\begin{align}
\left[\mathcal{I}^Q\right]_\vartheta = \sum_{j=1}^s\frac{n_j^2}{(1-\rho^2)}\left(\frac{u_{j_x}^2}{\sigma_{j_y}^2} + \frac{u_{j_y}^2}{\sigma_{j_x}^2} + \frac{2\rho u_{j_x} u_{j_y}}{\sigma_{j_x}\sigma_{j_y}}\right),
\label{eq:qfi_grid_rotation}
\end{align}
For consistency with the preceding subsection, we assume each source has the same standard deviation in the ordered basis of the grid, $\smash{\sigma_{j_x} = \sigma_{j_y} = \sigma}$. Then using the properties of the ceiling function and the modulo operation, we compute this sum to provide the following, more informative expression of the \textsc{qfi},
\begin{align}
\begin{split}
\left[\mathcal{I}^Q\right]_\vartheta &= \frac{s}{12\sigma^2(1-\rho^2)}\left\{d_x^2(N^2-1)[1 + \rho\sin(2\vartheta)]\right. \\
&+\left. d_y^2(M^2-1)[1 - \rho\sin(2\vartheta)]\right\}.
\label{eq:qfi_grid_rotation_2}
\end{split}
\end{align}
From this form, we observe two contributions to the \textsc{qfi} for arbitrarily chosen grid configurations \emph{and} source properties (specifically the correlation coefficient, $\rho$ and covariance matrix). The first provides a constant offset of the \textsc{qfi} and may be enhanced by increasing the number of emitters in the grid in either direction. Alternatively, for the same number of emitters, $s$, the \textsc{qfi} may be enhanced by increasing the mutual source separation distances. The second contribution provides an oscillatory dependence of the \textsc{qfi} on $\vartheta$, which emerges only when the following criteria are met
\begin{align}
\rho \neq 0, \quad \text{and,} \quad d_x^2 (N^2-1) \neq d_y^2 (M^2-1).
\label{eqn:osciallatory_requirement}
\end{align}
The first requirement is on the source properties and the second on the grid configuration. Provided both criteria are met, then from Eq.~\eqref{eq:qfi_grid_rotation_2} it is clear that the amplitude of the oscillatory behaviour may be enhanced by choosing a grid configuration with $N>M$ for positive $\rho$, and $N<M$ for negative $\rho$. The interplay between the requirements for a constant \textsc{qfi} can be clearly illustrated by the \textsc{qcrb}. From the inverse of Eq.~\eqref{eq:qfi_grid_rotation_2}, the \textsc{qcrb} is shown in Fig.~\ref{pic:qcrb_grid_rotations}. We note that if the requirements in Eq.~\eqref{eqn:osciallatory_requirement} are not met, the \textsc{qfi} can be made independent of the rotation angle $\vartheta$. The \textsc{qcrb} is then attainable without the use of adaptive strategies.

\section{Conclusions and Discussions} 
\label{sec:conclusions}

\noindent
In this work, we have developed a framework for detecting deformations applied to arbitrarily sized grid of sources. This finds many important practical applications in engineering. Our formalism allows the detection of stresses and strains subjected to materials, which provides the ability to prevent fractures before they eventuate. By tracking changes to the expected source positions during an applied deformation, our work is an analogue of the source localisation problem. This problem has recently received considerable research interest from the Laser Interferometer Gravitational-Wave Observatory (\textsc{ligo}) collaboration, who currently use triangulation results to localise signals radiating from compact binary coalescences. On the theoretical side, our formalism is the first to estimate the nature of deformations by use of quantum metrology, whilst maintaining full generality for unitary channels evolutions.  

Our approach uses the quantum Fisher information \textsc{qfi} as a figure of merit to estimate the type of deformation being administered to the grid. The sources comprising the undeformed grid are taken to be stationary, identical and independently distributed (\textsc{iid}) in a uniform manner with constant $x$-separation distance $d_x$, and $y$-separation $d_y$ between neighbouring emitters.  We model the $j$th source with a bi-normal spatial profile centred on $\smash{\bm{\mu}_j}$. Any deformation exacted affects only the expected source positions and not their covariances. This is physically motivated, since the nature of the sources should not be altered in the process. Then, the application of general deformations on the initial grid can be viewed as a unitary process, with grid dynamics described by the Hamiltonian $\smash{\hat{\mathcal{H}}}$. To ensure full generality, we allow the vector of parameters, $\smash{\bm{\varphi}}$, that fully describes the applied deformation, to appear as arbitrary parameterisations of both the eigenvalues and eigenvectors of $\smash{\hat{\mathcal{H}}}$. We solve the local generator of translations in $\smash{\bm{\varphi}}$, which is a generalisation of the result derived in~\cite{Pang2014_PRA} to multi-parameter quantum metrology. 

For the grid metrology formalism presented here, only the eigenvalues of the Hamiltonian depend on the vector of parameters $\smash{\bm{\varphi}}$, encoding information on the applied deformation. The generator describing dynamics in the parameters reduces to the set of commuting operators
\begin{align}
\hat{G}_j = \sum_{k = 1}^{n_g}\partial_j E_k\ket{\bm{n}}\bra{\bm{n}}.
\label{eqn:actual_generator_conclusions}
\end{align}
From this and the \textsc{iid} property of the sources---which permits each source to be treated in its individual Hilbert space---we derive the \textsc{qfim} from generator co-variances. Our expression of the \textsc{qfim} holds for any general grid deformation. We provide example applications to homogenous deformations, comprised of linear combinations of rotations, shearing, and, composite stretches. For composite grid stretches and shears, the deformation parameterises the Hamiltonian with the multiplicative factors $(\alpha, \beta)$. Since the factor stretch in each direction is independent, the \textsc{qfim} is diagonal. If a material is understood to be susceptible to fracture due to stretching along a specific direction, the sensitivity of \textsc{qfi} may be maximised by increasing the number of emitters along the direction in which the grid stretching eventuates, for an identical number of total emitters, $s$. The optimal grid configuration then becomes an array of sources, which is in agreement with results obtained in~\cite{Sidhu2017_PRA}, where the source optimisation problem was considered for enhancing estimates of source separation distances in stretched arrays. 

We also considered non-multiplicative Hamiltonian parameterisations resulting from grid rotations about any chosen axis. Oscillatory dependence of the \textsc{qfi}, that permit estimates on the amount of rotation, surfaces for particular grid configurations satisfying $\smash{d_x^2 (N^2-1) \neq d_y^2 (M^2-1)}$ and for non-zero correlation coefficient $\rho$ of the bi-normal spatial source distributions. 

A natural extension of the work considered here would be to consider the \textsc{qfi} in the far-field. The need for spatially propagating the field was circumvented here by working in the near field of the sources. This would motivate more realistic models and lead to the concept of the quantum Fisher information over a subset in space. Additionally, our formalism provides the machinery for non-commuting generators, multi-parameters, and, arbitrary parameterisations of the Hamiltonian. Since, the full extent of this formalism is not taken advantage of with quantum metrology of grid deformations, this would be fruitful future work.



\section{Acknowledgments}
\label{sec:acknowledgments}

\noindent
This research was funded by the DSTL Quantum 2.0 Technologies Programme.


\appendix
\section{Hermicity of generators}
\label{app:hermicity}

\noindent
The local generator of translations in $\varphi_j$ was found in Eq~\eqref{eqn:local_generator_integral_eq} to have the form 
\begin{align}
\hat{G}_j = \int_0^1 d\alpha \exp\left[-i\alpha\hat{\mathcal{H}}\right] \partial_j\hat{\mathcal{H}} \exp\left[i\alpha\hat{\mathcal{H}}\right].
\label{eqn:local_generator_integral_reminder}
\end{align}
The Hermitian conjugate of the operator requires knowledge of $\smash{(\partial_j\hat{\mathcal{H}})^\dagger}$. Since $\smash{\hat{G}_j}$ governs \emph{local} dynamics in $\varphi_j$ we write
\begin{align}
i\partial_j \ket{\psi} = \hat{G}_j\ket{\psi}.
\label{eqn:schrodinger_eqn}
\end{align}
Spectrally decomposing $\smash{\hat{H}}$ using the complete ortho-normal basis $\{p_j, \ket{\smash{\psi_j}}\}$, differentiating with respect to $\varphi_j$ and using Eq.~\eqref{eqn:schrodinger_eqn} we recover the Von Neumann equation
\begin{align}
\frac{d\hat{\mathcal{H}}}{d\varphi_j} = \partial_j\hat{\mathcal{H}} -i\left[\hat{G}_j, \hat{\mathcal{H}}\right].
\label{eqn:hamiltonian_dynamics}
\end{align}
The first term is the eigenvalue dependence on the parameter and is Hermitian. Since $\smash{(d_j\hat{\mathcal{H}})^\dagger = d_j\hat{\mathcal{H}}}$ then it follows that $\hat{G}_j^\dagger = \hat{G}_j$.


\section{Derivation of generator}
\label{sec:generator_derivation}

\noindent
The integral operator equation for the local generator of translations in $\varphi_j$ in Eq.~\eqref{eqn:local_generator_integral_eq} has been solved in~\cite{Wilcox1967_JMP, Pang2014_PRA}. We generalise the result for multi-parameter estimation variables. Defining the integrand as $\smash{\hat{Y}_m(\beta)}$, then we have the first order differential equation
\begin{align}
\partial_\beta\hat{Y}_m(\beta) = -i\left[\hat{\mathcal{H}}, \hat{Y}_m(\beta)\right] = -i\hat{\mathscr{H}}\left[\hat{Y}_m(\beta)\right],
\label{eqn:differential_operator_eqn}
\end{align}
where $\smash{\hat{\mathscr{H}}}$ is the Hermitian superoperator of $\hat{\mathcal{H}}$ defined by $\smash{\hat{\mathscr{H}}[\hat{O}] = [\hat{\mathcal{H}}, \hat{O}]}$, and the initial condition is $\smash{\hat{Y}_m(0) = \partial_m\hat{\mathcal{H}}}$. In Eq.~\eqref{eqn:normal_hamiltonian}, we defined the Hamiltonian $\hat{\mathcal{H}}$ to have $n_g$ unique eigenvalues $E_j$, $j\in[1, n_g]$, with degeneracies $d_j$, and corresponding eigenvectors $\ket{\smash{E_j^{(k)}}}$, $k \in [1, d_j]$. Hence, we write the superoperator $\smash{\hat{\mathscr{H}}}$ as 
\begin{align}
\hat{\mathscr{H}}(\bm{\varphi}) = \sum_{k,l=1}^{n_g} \sum_{i = 1}^{d_k}\sum_{j = 1}^{d_l} \lambda_{kl}^{(ij)}(\bm{\varphi}) \hat{\Gamma}_{kl}^{(ij)}(\bm{\varphi}),
\label{eqn:superop_spectrum}
\end{align}
with 
\begin{align}
\begin{split}
\lambda_{kl}^{(ij)}(\bm{\varphi}) &= E_k(\bm{\varphi}) - E_l(\bm{\varphi}), \\
\hat{\Gamma}_{kl}^{(ij)}(\bm{\varphi}) &= \ket{E_k^{(i)}(\bm{\varphi})}\bra{E_l^{(j)}(\bm{\varphi})},
\end{split}
\label{eqn:superop_evalue_vector}
\end{align}
and where the projectors satisfy $\smash{\hat{\Gamma}_{kl}^{(\alpha\beta)}\hat{\Gamma}_{mn}^{(\gamma\delta)} = \delta_{lm}\delta_{\beta\gamma}\hat{\Gamma}_{kn}^{(\alpha\delta)}}$. We can now solve the first order operator differential Eq.~\eqref{eqn:differential_operator_eqn} in this basis by writing $\smash{\hat{\mathscr{H}}[\hat{Y}_m(\beta)] = \lambda_{kl}^{(ij)}\hat{Y}_m(\beta)}$ and using the initial condition 
\begin{align}
\hat{Y}_m(0) = \partial_m\hat{\mathcal{H}} = \sum_{k,l}^{n_g}\sum_{i}^{d_k}\sum_{j}^{d_l} \Tr{\hat{\Gamma}_{kl}^{(ij)\dagger}\partial_m\hat{\mathcal{H}}}\hat{\Gamma}_{kl}^{(ij)},
\label{eqn:diff_eqn_initial_condition}
\end{align}
to obtain
\begin{align}
\hat{Y}_m(\beta) = \sum_{k,l}^{n_g}\sum_{i}^{d_k}\sum_{j}^{d_l} \Tr{\hat{\Gamma}_{kl}^{(ij)\dagger}\partial_m\hat{\mathcal{H}}}\hat{\Gamma}_{kl}^{(ij)} \exp\left[-\lambda_{kl}^{(ij)}\beta\right].
\label{eqn:diff_eqn_initial_condition}
\end{align}
Note that the zero eigenvalues of the superoperator occur when $k=l$ with degeneracy $\smash{r = \sum_jd_j^2}$. From the definition of $\smash{\hat{\mathcal{H}}}$ in Eq.~\eqref{eqn:normal_hamiltonian} and the properties of the projectors $\hat{\Gamma}_{kl}^{(ij)}$,
\begin{align}
  \Tr{\hat{\Gamma}_{kl}^{(ij)\dagger}\partial_m\hat{\mathcal{H}}} =
    \begin{cases}
      (\partial_mE_k)\delta_{ij} & \text{for $k=l$}\\
      (E_k - E_l)\braket{E_l^{(j)}\vert\partial_mE_k^{(i)}} & \text{for $k\neq l$}
    \end{cases}.
\end{align}
Substituting Eq.~\eqref{eqn:diff_eqn_initial_condition} for the integrand of Eq~\eqref{eqn:final_generator_general_form} and conducting the integration, we obtain the form of the local generator in the main body of the text.


\section{Alternative approach for deriving generator}
\label{sec:bch_approach_generator}

\noindent
In this section we use the Baker Campbell Hausdorff (\textsc{bch}) identity to solve the integral operator equation for $\smash{\hat{G}_j}$. Note that this approach is generally better suited for perturbative solutions for infinite series. The \textsc{bch} for a Hermitian operator $\smash{\hat{B}}$ and arbitrary operator $\smash{\hat{A}}$ reads
\begin{align}
\begin{split}
\exp\left[\mu\hat{B}\right]\hat{A}\exp\left[-\mu\hat{B}\right] &= \hat{A} + \mu\left[\hat{B}, \hat{A}\right] + \frac{\mu^2}{2!}\left[\hat{B}, \left[\hat{B}, \hat{A}\right]\right] + \ldots, \\
&= \sum_{n=0}^\infty\frac{\mu^n}{n!}C^n_{\hat{B}}(\hat{A}),
\label{eqn:bch_identity}
\end{split}
\end{align}
for some $\smash{\mu \in \mathbbm{R}}$ and where we have defined $\smash{C^n_{\hat{B}}(\hat{A})}$ as the $n$th--order nested commutator of $\smash{\hat{A}}$ and $\smash{\hat{B}}$ (note the subscript $n$ \emph{does not} refer to a power). Using Eq.~\eqref{eqn:bch_identity} to re-express the integrand of Eq.~\eqref{eqn:local_generator_integral_eq} and performing the integration, we obtain
\begin{align}
\begin{split}
\hat{G}_j &= \hat{\mathcal{H}}_j - \frac{i}{2!} \left[\hat{\mathcal{H}}, \hat{\mathcal{H}}_j\right] - \frac{1}{3!}\left[\hat{\mathcal{H}}, \left[\hat{\mathcal{H}}, \hat{\mathcal{H}}_j\right]\right] + \ldots, \\
&= \sum_{n=0}^\infty \frac{(-i)^n}{(n+1)!}C^n_{\hat{\mathcal{H}}}\left(\hat{\mathcal{H}}_j\right), \\
&= g\left[-iC_{\hat{\mathcal{H}}}\right]\left(\hat{\mathcal{H}}_j\right),
\label{eqn:generator_from_bch_identity}
\end{split}
\end{align}
where $\smash{\hat{\mathcal{H}}_j = \partial_j\hat{\mathcal{H}}}$ and where we defined the generating function of the expansion coefficients in Eq.~\eqref{eqn:generator_from_bch_identity} as
\begin{align}
g[t] = \frac{\exp\left[t\right] - 1}{t}.
\label{eqn:generating_function}
\end{align}
Eq.~\eqref{eqn:generator_from_bch_identity} is a series solution for the local generator of translations for the parameter $\smash{[\bm{\varphi}]_j}$. It reproduces the adjoint action series of Duhamel's formula presented in the main body of the text. To detail how, we consider the same spectral decomposition of the Hamiltonian, Eq.~\eqref{eqn:normal_hamiltonian}, used in solving Duhamel's formula. Then we have
\begin{align}
\hat{\mathcal{H}}_j = \sum_{k=1}^{n_g} \sum_{l=1}^{d_j} \left[\partial_j E_k\ket{E_k^{(l)}}\bra{E_k^{(l)}} + E_k \partial_j \left(\ket{E_k^{(l)}}\bra{E_k^{(l)}}\right)\right].
\label{eqn:normal_hamiltonian_derivative_appendix}
\end{align}
Using the resolution of the identity and the orthonormality criterion of the projectors of the Hamiltonian, the $n$th--order nested commutator of $\smash{\hat{\mathcal{H}}}$ and $\smash{\hat{\mathcal{H}}_j}$ becomes
\begin{align}
\begin{split}
C^n_{\hat{\mathcal{H}}} &= -\sum_{a,b,c,d}(E_a- E_c)^{n+1}\ket{E_a^{(b)}}\Braket{E_a^{(b)}\middle\vert\partial_j E_c^{(d)}}\bra{E_c^{(d)}},\\
&= -\sum_{a,b,c,d}(E_a- E_c)^{n+1}\hat{\mathcal{A}}_{a,b,c,d},
\label{eqn:nested_commutators}
\end{split}
\end{align}
for $\smash{n\geq1}$, where we have collected all summations under the same sign for brevity, although they are paired such that $a, c$ sum of distinct eigenvalues and $b, d$ sum over the eigenvector degeneracies for eigenvalues $\smash{E_a}$, $\smash{E_c}$ respectively, and, where we introduce the weighted projector 
\begin{align}
\hat{\mathcal{A}}_{a,b,c,d} = \Braket{E_a^{(b)}\middle\vert\partial_j E_c^{(d)}}\hat{\Gamma}_{ac}^{(bd)},
\label{eqn:projector_definition}
\end{align}
for brevity. Clearly then the $n$th-order nested commutators captures the dynamics resulting from the eigenvectors dependence on $\smash{\bm{\varphi}}$. The zeroth-order nested commutator is, by definition, the derivative of the Hamiltonian, written in Eq.~\eqref{eqn:normal_hamiltonian_derivative_appendix}. By inspection the first term of $\smash{\hat{\mathcal{H}}_j}$ reproduces the eigenvalue contribution to the local generator of translations $\smash{\hat{G}_j}$ written in the main body of the text. By use of the resolution of the identity, the eigenvector dependence of $\smash{\hat{\mathcal{H}}_j}$ (second term) is in fact the result obtained by $\smash{C^0_{\hat{\mathcal{H}}}}$, such that
\begin{align}
\hat{\mathcal{H}}_j = \sum_{k=1}^{n_g} \partial_j E_k\hat{P}_k + C^0_{\hat{\mathcal{H}}}.
\label{eqn:hamiltonian_derivative_appendix_2}
\end{align}
Our task now is to reproduce the eigenvector dependence of the generator, so we omit the first term of $\smash{\hat{\mathcal{H}}_j}$. Then, combining the result of the $n$th-order nested commutators in Eq.~\eqref{eqn:nested_commutators} and Eq.~\eqref{eqn:hamiltonian_derivative_appendix_2} with the series solution of the generator in Eq.~\eqref{eqn:generator_from_bch_identity}, we have
\begin{align}
\begin{split}
\hat{G}_j &\propto  i\sum_{a,b,c,d}\left(\sum_{n=0}^\infty \frac{[-i(E_a-E_c)]^n}{n!} - 1\right)\mathcal{A}(a,b,c,d),\\
&\propto i\sum_{a,b,c,d}\left(\exp[-i(E_a-E_c)] - 1\right)\mathcal{A}(a,b,c,d),
\label{eqn:final_generator_from_bch_identity}
\end{split}
\end{align}
where we write the proportionality to indicate our omission of the term with eigenvalue dependence. By use of the identity
\begin{align}
\exp[-i(E_a-E_c)] - 1 = -2i\exp\left[-\frac{i(E_a-E_c)}{2}\right]\sin\left[\frac{(E_a-E_c)}{2}\right],
\label{eqn:complex_sin_identity}
\end{align}
we arrive at the same form of the local generator of translations $\smash{\hat{G}_j}$ as presented in the main body of the text.

\bibliographystyle{apsrev4-1}

\end{document}